\documentclass[10pt,a4paper]{article}

\usepackage[a4paper,margin=1in]{geometry}
\usepackage{amsmath,amsthm}
\usepackage{newtxtext}
\usepackage[subscriptcorrection]{newtxmath}
\usepackage{graphicx}
\usepackage{booktabs}
\usepackage[round]{natbib}
\usepackage{authblk}
\usepackage[plain,noend]{algorithm2e}
\usepackage[colorlinks=true,linkcolor=blue!70!black,citecolor=blue!70!black,urlcolor=blue!70!black]{hyperref}
\makeatletter
\AtBeginDocument{\def\NR@gettitle#1{\def\@currentlabelname{}}}
\makeatother

\theoremstyle{plain}
\newtheorem{theorem}{Theorem}
\newtheorem{proposition}{Proposition}
\newtheorem{lemma}{Lemma}

\theoremstyle{definition}

\newcommand{\tbl}[2]{\caption{#1}\par\smallskip\centering#2}
\newenvironment{tabnote}{\par\vspace{4pt}\footnotesize\noindent}{\par}
\newcommand{\figuresize}[1]{\gdef\figscale{#1}}
\figuresize{1}
\makeatletter
\def\figurebox#1#2#3[#4]{\centering\includegraphics[width=\figscale\linewidth]{#4}}
\makeatother

\def\T{{\mathrm{\scriptscriptstyle T}}}
\def\var{\mathrm{var}}

\def\pr{\mathrm{pr}}
\def\sign{\mathrm{sign}}
\def\daw{\mathrm{daw}}

\newcommand{\E}{E}
\newcommand{\Pp}{\mathrm{pr}}
\newcommand{\R}{\mathbb R}
\newcommand{\ind}{1}

\newcommand{\Var}{\mathrm{var}}
\newcommand{\Cov}{\mathrm{cov}}

\newcommand{\sgn}{\mathrm{sign}}
\newcommand{\op}{o_p}
\newcommand{\Op}{O_p}
\newcommand{\dto}{\to}
\newcommand{\pto}{\to}
\newcommand{\N}{N}
\newcommand{\calT}{\mathcal T}
\newcommand{\trans}{^{\T}}
\newcommand{\cG}{\mathcal{G}}

\newcommand{\Nref}{N}

\newenvironment{keywords}{\vspace{6pt}\noindent\textbf{Keywords:}\ }{\par\vspace{12pt}}

\title{\textbf{Non-normal Limits and Joint Calibration of Cross-fitted Heterogeneity Tests}}

\author[1]{Even He\thanks{Email: \texttt{heyiwen5@gmail.com}}}
\author[1]{Zaizai Yan\thanks{Email: \texttt{zz.yan@163.com}}}
\affil[1]{College of Science, Inner Mongolia University of Technology, Hohhot 010051, China}

\date{}

\begin{document}

\maketitle

\begin{abstract}
Cross-fitting is used to test whether a ranking learned in a randomized experiment separates units that benefit from treatment from those that do not: each fold is ranked by a model fitted to the other folds, and the standardized difference in estimated effects between units ranked high and low is compared with a normal distribution. Under the null hypothesis of no heterogeneity the ranking is determined by noise, and because each fold helps to rank the others the fold estimates are dependent. We show that this dependence changes the shape of the null distribution, not just its variance. For least squares on one covariate the limit is bimodal for two folds and skewed for many, no variance correction restores the level, and the model is a counterexample to a published central limit theorem. With many covariates the limit approaches normality, but its variance stays inflated. Removing the non-normal component of the statistic removes its power, so we keep the statistic and change its reference distribution: joint calibration simulates together the moments that choose the rankings and those that evaluate them and applies the same selection to the draws; it has asymptotically exact level and requires no refitting. Applications to clinical and marketing trials, together with random treatment reassignments on their covariates, demonstrate that the standard test substantially inflates Type I error and can produce spurious rejections that joint calibration avoids.
\end{abstract}


\section{Introduction}

Randomized experiments are increasingly analysed with rankings learned by machine learning: a model of the conditional average treatment effect, such as a causal forest or a metalearner \citep{athey2016recursive,wager2018estimation,kunzel2019metalearners,nie2021quasi}, is fitted to part of the data and used to order the remaining units by predicted benefit. Sorted group average treatment effects \citep{chernozhukov2025generic,imai2025} ask whether such a ranking separates units that benefit from units that do not, by comparing the average effects of the units ranked above and below the median. To avoid evaluating a ranking on the data that produced it, the data are split \citep{cox1975note}: the sample is divided into folds, the ranking applied to each fold is learned from other folds, and the fold estimates are averaged. The standardized average is then compared with a normal distribution.

For a fixed target this comparison is justified. With orthogonal scores, the estimate from each fold is asymptotically linear in an influence function that does not depend on the training data, so the fold estimates are asymptotically independent and their average is asymptotically normal \citep{chernozhukov2018dml,chernozhukov2023simple}. The null hypothesis of no heterogeneity breaks the argument. If every unit has the same conditional average treatment effect, every ranking is optimal, and the value of the optimal ranking is not a differentiable functional of the distribution of the data \citep{hirano2012impossibility,luedtke2016nonunique}; the test error of a learned classifier has the same property \citep{laber2011adaptive}. A consistent learner converges to a constant function, but its estimation error still decides which half of a fold is called the high-effect group, so the learned ranking is determined by noise. In the usual design, in which each fold is ranked by a model fitted to all the other folds, the noise of one fold helps to choose the ranking applied to another, and conversely. Each fold estimate still has mean zero given its ranking, but the fold estimates are no longer independent.

We show that this dependence changes the shape of the limiting distribution, not only its variance. The analysis does not rest on the causal setting: it applies whenever some folds choose, through a fixed number of their sample moments, the weights with which another fold is evaluated, and the evaluation has mean zero under the null hypothesis whatever weights are chosen. We call the moments of a fold that help to choose rankings its training moments, and those to which a ranking is applied its evaluation moments. Conditionally on the training moments, the limit is normal. Unconditionally it is the sum of a normal variable and a non-normal one, weighted by the correlation between the training and evaluation moments of a fold; it is normal when this correlation is zero and, in the linear models that we study, only then. The non-normal component cannot simply be removed, because it is the component that a signal shifts: a statistic without it has a normal limit but no power against the alternatives that the learner is designed to detect.

For least squares on one covariate the limit is explicit. With two folds its density has a closed form and is bimodal, with many folds it is skewed, and standardizing by the exact variance does not restore the nominal level; the best linear predictor test, in which the two moments coincide, rejects almost twice as often as it should with two folds. This model satisfies the conditions of Theorem~6 of \citet{imai2025}, which asserts asymptotic normality of the standardized group effect, and is therefore a counterexample to it. With many covariates the limit approaches normality while its variance stays inflated by a factor that does not decrease with the number of folds. The variance corrections of the cross-validation literature err in either direction, depending on the numbers of covariates and folds, whereas a correction estimated from the training and evaluation moments restores the level when the covariates are many. Under local alternatives the limit stays non-normal, and the strength of signal at which normality returns cannot be estimated consistently. The same structure yields a test, joint calibration: a parametric bootstrap that draws the training and evaluation moments jointly from their estimated normal distribution and applies the selection to the draws. Its level is asymptotically exact, whichever folds train each ranking, and it requires no refitting of the learner.

The dependence of cross-validation folds is well known. Resampled $t$-tests over-reject \citep{dietterich1998approximate}, corrections based on moments allow for training sets that overlap \citep{nadeau2003inference,markatou2005analysis}, the variance of $K$-fold cross-validation has no estimator that is unbiased under every distribution \citep{bengio2004variance}, nested cross-validation repairs intervals for prediction error \citep{bates2024cv}, and central limit theorems for cross-validated risk hold under stability conditions \citep{austern2020,bayle2020cv}, which even two stable learners can violate when they are compared \citep{bayle2026relative}. These results concern the variance of a cross-validated estimate; here the shape changes as well, and the stability they assume fails under the null hypothesis of no heterogeneity. \citet{fava2025training} proves a central limit theorem for averages over splits that requires only convergence of the learned model, and notes that normal intervals can fail when the model converges to a constant, as it does here; we describe the limit in that case and calibrate the test.

A single split gives valid inference for the target it selects, and conditional separations can calibrate post-reduction confidence sets \citep{rinaldo2019bootstrapping,rasines2023splitting,battey2026post}; $p$-values from many splits can also be combined \citep{meinshausen2009pvalues,diciccio2020exact}, whereas cross-fitting uses every fold in both roles. Tests of heterogeneity that learn no ranking \citep{crump2008nonparametric,ding2016randomization} and evaluations of a ranking on held-out data \citep{yadlowsky2025rate} avoid the dependence. \citet{kim2024crossu} found a non-normal limit when two halves of a sample swap the roles of choosing a direction and evaluating it, and \citet{guo2025rts} used this example to motivate rank-transformed subsampling; after rescaling, that limit is the member of our family in which the training and evaluation moments are perfectly correlated. \citet{imai2025comment} recommend the cross-fitted estimator of group effects with a variance that accounts for training and evaluating on the same data, through a lemma of \citet{nadeau2003inference}, and a normal interval. Under the null hypothesis of no heterogeneity and with few covariates, our results imply that such an interval is not valid even with the exact variance. \citet{wager2025sequential} evaluates each fold only with rankings trained on earlier folds; we keep the usual statistic and replace its reference distribution.

\section{Cross-fitted tests and their limit}\label{sec:setup}\label{sec:general}

\subsection{Setting and the usual test}\label{sec:usual}

We observe independent copies $(X_i,A_i,Y_i)$ $(i=1,\ldots,n)$ of covariates $X$, a binary treatment $A$ assigned with known probability $p$ independently of $X$, and an outcome $Y=AY(1)+(1-A)Y(0)$, with no interference between units. Write $\tau(x)=E\{Y(1)-Y(0)\mid X=x\}$ for the conditional average treatment effect. The transformed outcome $\tilde Y=AY/p-(1-A)Y/(1-p)$ satisfies $E(\tilde Y\mid X)=\tau(X)$. The null hypothesis of no heterogeneity states that $\tau(X)=\tau$ almost surely for an unknown constant $\tau$; individual effects may still vary.

The units are divided at random into $K$ folds $I_1,\ldots,I_K$ of size $m$, and $P_kv=m^{-1}\sum_{i\in I_k}v_i$ denotes an average over fold $k$. For each fold $k$, a set of training folds $F(k)\subseteq\{1,\ldots,K\}\setminus\{k\}$ supplies the treatments and outcomes from which a score $s_k$ is learned; covariates of any fold may be used. The usual choice, leave-one-fold-out, is $F(k)=\{1,\ldots,K\}\setminus\{k\}$. Fold $k$ is divided at the median of $s_k$ over the fold: $\hat g_k(x)=1$ if $s_k(x)$ lies above that median and $\hat g_k(x)=-1$ otherwise. The estimated difference between the average effects of the two halves of fold $k$, and its average over folds, are
\[
D_k=2P_k(\hat g_k\tilde Y),\qquad D_n=K^{-1}(D_1+\cdots+D_K).
\]
The two halves of a fold have equal size, so their estimated average effects are $P_k\tilde Y\pm D_k/2$. With $Q_n=n^{-1}(\tilde Y_1+\cdots+\tilde Y_n)$, the estimated average effect in the upper halves is therefore $\hat\theta_+=Q_n+D_n/2$. These are the estimators of \citet{imai2025}.

The usual test treats the $n$ contributions $2\hat g_k(X_i)\tilde Y_i$ $(i\in I_k;\ k=1,\ldots,K)$ as independent and divides $D_n$ by the resulting standard error $s_n$ of their mean, which agrees to first order with the within-fold variance estimator of \citet{bayle2020cv}. Under the null hypothesis each $D_k$ has mean zero given its ranking, because $\hat g_k$ is computed without the treatments and outcomes of fold $k$ and sums to zero over the fold (Lemma~\ref{lem:center} of the Supplementary Material). Hence $s_n^2$ estimates the sum of the variances of the fold estimates, divided by $K^2$, and omits their covariances.

\subsection{Training and evaluation moments}\label{sec:moments}

To see why the limit need not be normal, separate the moments that choose a ranking from those that evaluate it. Take first the simplest learner, least squares of $\tilde Y$ on one covariate. Fold $k$ is ranked by $x$ times the sign of the slope fitted on its training folds, and because the denominator of the slope is positive, that sign is the sign of $\sum_{r\in F(k)}P_r\{(x-P_rx)\tilde Y\}$. Each training fold $r$ thus contributes a single number, its training moment $P_r\{(x-P_rx)\tilde Y\}$, and its treatments and outcomes reach the ranking in no other way. Fold $k$ is in turn evaluated through a single number, its evaluation moment $P_k(g\tilde Y)$, where $g$ is the median split of $x$. Every fold therefore plays two roles, each through one moment, and the two moments of a fold are correlated because they multiply the same outcomes.

The general setting keeps this structure and has three ingredients. The first is a feature map $h:\mathcal X\to\mathbb R^d$, whose moments train the rankings. The second is a set of $J$ candidate rankings, each represented by a bounded weight $g_j$ with mean zero under the distribution of $X$; the moments of these weights evaluate a fold. Median splits on single covariates are an example, and contrasts between the top and bottom quintile groups of a covariate, as in \citet{chernozhukov2025generic}, are another once the quantile conditions of \S~\ref{sec:quantile} are imposed at each threshold. The third is a selection function $\omega:\mathbb R^d\to\mathbb R^J$, bounded and unchanged by positive rescaling of its argument, which turns the summed training moments into a weight on each candidate; for instance, $\omega$ may select the candidate with the largest absolute training moment, with its sign. The example above has $h(x)=x$, $J=1$, $g_1(x)=\sign(x)$ and $\omega(z)=\sign(z)$.

Let $\nu=\tilde Y-\tau$,
\[
V=\bigl[\{h(X)-E h(X)\}^\T\nu,\ g_1(X)\nu,\ \ldots,\ g_J(X)\nu\bigr]^\T,\qquad \Sigma=E(VV^\T),
\]
and partition $\Sigma$ into blocks $\Sigma_{ZZ}$, $\Sigma_{ZU}=\Sigma_{UZ}^\T$ and $\Sigma_{UU}$ of sizes $d$ and $J$. Under the null hypothesis $\Sigma_{ZZ}$ is the limiting covariance matrix of the training moments of a fold, $\Sigma_{UU}$ that of its evaluation moments, and $\Sigma_{ZU}$ the covariance between the two; the last block decides whether the limit is normal. In fold $k$ the training moment and the evaluation moments are
\[
Z_{k,n}=m^{1/2}P_k\{(h-P_kh)\tilde Y\},\qquad U_{k,n}=m^{1/2}\{P_k(\hat g_{k1}\tilde Y),\ldots,P_k(\hat g_{kJ}\tilde Y)\}^\T,
\]
where $\hat g_{kj}$ is $g_j$ computed with the empirical median of fold $k$ in place of the population median, so that it sums to zero over the fold and $\tau$ drops out. The ranking applied to fold $k$ is chosen from the training moments of the folds that train it, $\hat\omega_k=\omega(\sum_{r\in F(k)}Z_{r,n})$. With an estimate $\hat\Sigma_{UU}$ of $\Sigma_{UU}$, the statistic is
\begin{equation}\label{eq:Tn}
T_n=(K\hat v)^{-1/2}\sum_{k=1}^K\hat\omega_k^\T U_{k,n},\qquad \hat v=K^{-1}\sum_{k=1}^K\hat\omega_k^\T\hat\Sigma_{UU}\hat\omega_k,
\end{equation}
which to first order is the usual standardized estimate of \S~\ref{sec:usual}. Two familiar cases lie at the ends of this setting: when the training and evaluation moments are uncorrelated the usual normal theory holds (Proposition~\ref{prop:structure}), and when they are perfectly correlated the two-fold limit is the one found by \citet{kim2024crossu} for a cross U-statistic (\S~\ref{sec:univariate}).

\subsection{A general limit}\label{sec:generallimit}

\begin{theorem}\label{thm:general}
Suppose that the null hypothesis holds, that $E[\{1+\|h(X)\|^2\}(1+\nu^2)]<\infty$, that $K$, $d$ and $J$ are fixed, that the quantile conditions of \S~\ref{sec:quantile} hold, and that $\hat\Sigma_{UU}$ converges in probability to $\Sigma_{UU}$. Let $(Z_k^\T,U_k^\T)^\T$ $(k=1,\ldots,K)$ be independent $N(0,\Sigma)$ vectors, $\omega_k=\omega(\sum_{r\in F(k)}Z_r)$ and $v=K^{-1}\sum_k\omega_k^\T\Sigma_{UU}\omega_k$. If, with probability one, $v>0$ and $\omega$ is continuous at each $\sum_{r\in F(k)}Z_r$, then $T_n$ converges in distribution to
\begin{equation}\label{eq:TSigma}
T_\Sigma=(Kv)^{-1/2}\sum_{k=1}^K\omega_k^\T U_k.
\end{equation}
\end{theorem}

The proof in \S~\ref{app:calibproof} combines a joint central limit theorem for the moments of all folds with the continuous mapping theorem. The limit keeps the selection step: the Gaussian vectors pass through the same function $\omega$ as the data. Theorem~\ref{thm:general} therefore covers every ranking that is chosen from finitely many candidates and depends on the treatments and outcomes of its training folds only through a fixed number of moments: least squares on one covariate, a choice among the median splits of several covariates by their training moments, and combinations of the two. Least squares on several covariates chooses from a continuum of rankings and is treated in \S~\ref{sec:directions}. Nor is the theorem specific to treatment effects. The null hypothesis enters only through $E(\nu\mid X)=0$, and the treatment only through the transformed outcome, so the same limit governs any cross-fitted statistic in which training folds choose, through a fixed number of moments, the weights with which another fold is evaluated, provided that the evaluation has mean zero under the null hypothesis whatever weights are chosen. The cross U-statistic of \citet{kim2024crossu} is one instance outside the causal setting.

\subsection{Where non-normality comes from}\label{sec:structure}

\begin{proposition}\label{prop:structure}
Suppose that $\Sigma_{ZZ}$ is invertible and let $\Gamma=\Sigma_{UZ}\Sigma_{ZZ}^{-1}$. Conditionally on $Z_1,\ldots,Z_K$, the variable $T_\Sigma$ in \eqref{eq:TSigma} is normal with mean and variance
\[
\mu=(Kv)^{-1/2}\sum_{k=1}^K\omega_k^\T\Gamma Z_k,\qquad 1-(Kv)^{-1}\sum_{k=1}^K\omega_k^\T\Gamma\Sigma_{ZU}\,\omega_k .
\]
In particular, $T_\Sigma$ has the standard normal distribution if $\Sigma_{ZU}=0$.
\end{proposition}

\begin{proof}
Given the training moments, $U_k=\Gamma Z_k+E_k$, where $E_1,\ldots,E_K$ are independent $N(0,\Sigma_{UU}-\Gamma\Sigma_{ZU})$ vectors independent of $Z_1,\ldots,Z_K$, and $\omega_1,\ldots,\omega_K$ and $v$ are functions of $Z_1,\ldots,Z_K$. The conditional mean and variance follow, and they equal $0$ and $1$ when $\Sigma_{ZU}=0$.
\end{proof}

Proposition~\ref{prop:structure} locates the non-normality. Given the training moments, the part of the evaluation moments that is uncorrelated with them contributes a normal variable, whatever rankings are chosen. The remainder, $\mu$, collects the part of each fold's evaluation moments that its own training moment predicts, weighted by the ranking that the other folds chose for it. Each $Z_k$ thus enters $T_\Sigma$ twice, through its own term in $\mu$ and through the rankings of the folds that fold $k$ trains, and integrating over the training moments gives a mixture of normal distributions whose means and variances depend on the rankings. Normality does not require the selection to converge: when $\Sigma_{ZU}=0$ the limit is normal even if the selected ranking never settles. In the models of \S~\ref{sec:linear} the mixture is non-normal whenever the correlation is non-zero.

The non-normal component cannot be removed without removing the power. Evaluating each fold through the residual $U_k-\Gamma Z_k$ instead of $U_k$ would give, after standardization, a standard normal limit for any selection, since this residual is independent of all training moments. Suppose, however, that $\var(\nu\mid X)$ is constant and that the alternative is local and linear in the features, $\tau(x)=\tau+m^{-1/2}\theta^\T h_0(x)$ with $h_0=h-E h(X)$. Then $U_k$ and $Z_k$ have means $E\{g(X)h_0(X)^\T\}\theta$ and $E\{h_0(X)h_0(X)^\T\}\theta$, where $g=(g_1,\ldots,g_J)^\T$, while $\Gamma=E\{g(X)h_0(X)^\T\}[E\{h_0(X)h_0(X)^\T\}]^{-1}$, so the residual has mean zero under the alternative as well: the modified statistic cannot detect the very alternatives that a learner using $h$ is designed to find. In simulations of the model of \S~\ref{sec:univariate} it rejected about 5\% of the time both under the null hypothesis and under two alternatives against which the usual statistic, with normal critical values, rejected 37\% and 89\% of the time (\S~\ref{app:gatessim}). Since a signal acts through the non-normal component, \S~\ref{sec:calibration} keeps the usual statistic and changes its reference distribution instead.

\section{Linear rankings}\label{sec:linear}

\subsection{One covariate}\label{sec:univariate}

This section makes the general limit explicit for rankings learned by least squares. We begin with one covariate, where the learner chooses only the sign of a slope, the limits form a one-parameter family, and their shape can be described exactly. The same model satisfies the conditions of a published central limit theorem and is a counterexample to it (\S~\ref{sec:published}). With several covariates (\S~\ref{sec:directions}) the learner chooses a direction, and the limit moves towards normality while its variance stays inflated.

Let $X$ and $\varepsilon$ be independent standard normal variables, $Y(1)=Y(0)=\varepsilon$, assign half of the units to treatment by complete randomization, and split each arm at random into $K$ folds. The ranking applied to fold $k$ orders units by $\hat\beta_{(k)}x$, where $\hat\beta_{(k)}$ is the least-squares slope, without intercept, of $\tilde Y=2(2A-1)\varepsilon$ on $X$ in the other $K-1$ folds. The learner is consistent, $\hat\beta_{(k)}=O_p(m^{-1/2})$, but its sign does not converge. Since the denominator of the slope is positive, its sign is that of the sum of the training moments of the other folds, and this is the case $h(x)=x$, $J=1$, $g_1(x)=\sign(x)$ and $\omega(z)=\sign(z)$ of \S~\ref{sec:moments}. The training moment of a fold averages $X\tilde Y$ and its evaluation moment averages $\sign(X)\tilde Y$, so their correlation is $\rho=E\{X\sign(X)\}=E|X|=(2/\pi)^{1/2}$. For $\rho\in[-1,1]$ define
\begin{equation}\label{eq:Trho}
T_K(\rho)=K^{-1/2}\sum_{k=1}^K\sign\Bigl(\sum_{r\neq k}Z_r\Bigr)U_k,
\end{equation}
where $(Z_k,U_k)$ $(k=1,\ldots,K)$ are independent pairs of standard normal variables with correlation $\rho$.

\begin{theorem}\label{thm:counter}
In this model, with leave-one-fold-out and $K\geq2$, every group effect of every learned ranking, and its average over training samples, is zero. Moreover, $n^{1/2}D_n/4$ converges in distribution to $T_K\{(2/\pi)^{1/2}\}$ and $n^{1/2}\hat\theta_+/2^{3/2}$ to $T_K(\pi^{-1/2})$, second moments converge, and
\begin{equation}\label{eq:varTK}
\var\{T_K(\rho)\}=1+\frac{2\rho^2(K-1)}{\pi(2K-3)^{1/2}}.
\end{equation}
For every $K\geq2$ and $\rho\neq0$, no rescaling of $T_K(\rho)$ is normal. In particular, $\hat\theta_+/\var(\hat\theta_+)^{1/2}$ does not converge in distribution to $N(0,1)$.
\end{theorem}

The scalings are those under which the statistics would have unit variance if the folds were independent, since $\tilde Y$ has variance 4. The effect in the upper half has the smaller correlation $\pi^{-1/2}$ because its weight $1+\sign(x)$ has second moment 2. The fold moments converge jointly to independent normal pairs with correlation $\rho$, and Theorem~\ref{thm:general} gives the limit. Two Gaussian integrations by parts show that two distinct terms of \eqref{eq:Trho} have covariance $2\rho^2/\{\pi(2K-3)^{1/2}\}$, where $2K-3=(K-1)^2-(K-2)^2$ is the determinant of the covariance matrix of the two sums of training moments; this gives \eqref{eq:varTK}. For non-normality, $T_K(\rho)$ is, given the signs, a combination of the $U_k$ with unit conditional variance, so $E\exp\{tT_K(\rho)\}\leq2^K\exp(t^2/2)$ for every $t$, a bound that any normal distribution with variance above one violates for large $t$. The details are in \S~\ref{app:limits}.

The next result describes the distribution of $T_K(\rho)$ and, for two folds, gives it in closed form.

\begin{proposition}\label{prop:shape}
Let $B_K=K^{-1/2}\sum_{k}\sign(\sum_{r\neq k}Z_r)Z_k$, where $Z_1,\ldots,Z_K$ are independent standard normal variables, and let $W$ be a standard normal variable independent of $B_K$.
\begin{enumerate}
\item For every $\rho\in[-1,1]$, $T_K(\rho)$ has the distribution of $\rho B_K+(1-\rho^2)^{1/2}W$.
\item For $K=2$, $B_2=\sign(Z_1Z_2)(|Z_1|+|Z_2|)/2^{1/2}$, the fourth cumulant of $T_2(\rho)$ is $-(4/\pi+12/\pi^2)\rho^4$, and
\begin{equation}\label{eq:cf2}
E\exp\{itT_2(\rho)\}=\exp(-t^2/2)-(4/\pi)\exp\{-(1-\rho^2)t^2/2\}\,\daw(\rho t/2)^2,
\end{equation}
where $\daw(x)=\exp(-x^2)\int_0^x\exp(y^2)\,{\rm d}y$ is Dawson's integral. For $|\rho|<1$ let $\sigma=(1-\rho^2)^{1/2}$, $a=|\rho|/\sigma$, $b=|\rho|/(1+\sigma^2)^{1/2}$ and $r=\sigma/(1+\sigma^2)^{1/2}$. Then $T_2(\rho)$ has density
\begin{equation}\label{eq:density2}
f_\rho(x)=2\phi(x)\{2\Phi_r(ax,bx)+2\Phi_r(-ax,-bx)-1\},
\end{equation}
where $\phi$ is the standard normal density and $\Phi_r$ is the bivariate standard normal distribution function with correlation $r$.
\item For $K=2$ and $|\rho|<1$, $f_\rho''(0)=(4/\pi)\phi(0)\{\rho^2/(\sigma+\sigma^3)-\arctan\sigma\}$. Hence $x=0$ is a strict local minimum of $f_\rho$ if $|\rho|>\rho_*$ and a strict local maximum if $|\rho|<\rho_*$, where $\rho_*=0.748$ is the unique root of $\rho^2=(\sigma+\sigma^3)\arctan\sigma$.
\end{enumerate}
\end{proposition}

\begin{figure}
\figuresize{.85}
\figurebox{20pc}{25pc}{}[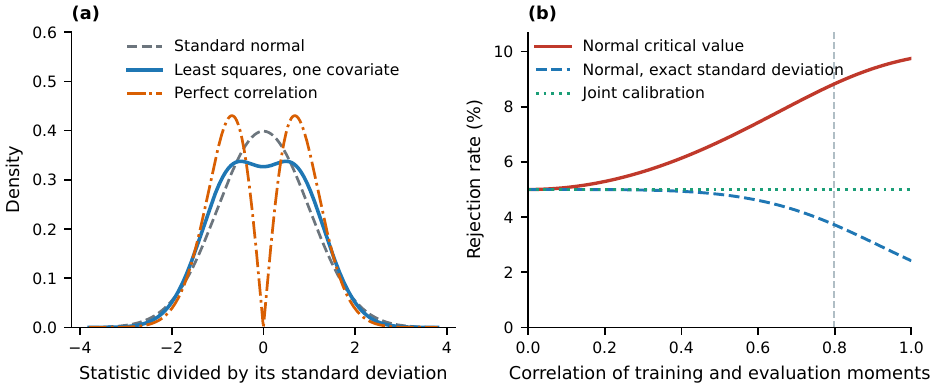]
\caption{The two-fold limit with one covariate moves away from the normal distribution as the correlation $\rho$ between the training and evaluation moments grows. (a) Density of the limit, scaled to unit variance. (b) Rejection rate of a nominal 5\protect\% two-sided test; the vertical line marks least squares on a normal covariate.}
\par\smallskip{\footnotesize Alt text: Two line plots. Panel (a) shows three densities on a common scale: the standard normal, the two-fold limit for least squares on one covariate, which is lower and flatter at zero with two shallow peaks near $\pm0.6$, and the limit under perfect correlation, which has two peaks near $\pm0.7$ and falls to zero at the origin. Panel (b) shows rejection rates against the correlation from 0 to 1: with normal critical values the rate rises from 5\% to nearly 10\%, with the exact standard deviation it falls from 5\% to about 2.5\%, and joint calibration stays at 5\%.}
\label{fig:limit}
\end{figure}

Part~(i) follows from Proposition~\ref{prop:structure}, and the proofs of (ii) and (iii) are in the Appendix. Part~(i) separates two components. The part of the evaluation moments that is uncorrelated with the training moments contributes the exactly normal variable $W$; all departure from normality comes from $B_K$, whose distribution depends only on the number of folds, and $\rho$ is its weight in the statistic. The characteristic function \eqref{eq:cf2} takes negative values, for example $-0.039$ at $t=2$ when $\rho=(2/\pi)^{1/2}$, so $T_2(\rho)$ is not a scale mixture of centred normal distributions, whose characteristic functions are positive. By (iii), least squares on a normal covariate, with $\rho=0.798$, has a bimodal two-fold limit with modes at $\pm0.59$, whereas the estimated effect in the upper half, with $\rho=\pi^{-1/2}=0.564$, has a unimodal one (Fig.~\ref{fig:limit}(a)). Evaluated numerically on a fine grid of $\rho$, $f_\rho$ has two modes when $|\rho|>\rho_*$ and one otherwise. With $K=2$ and $\rho=(2/\pi)^{1/2}$, normal critical values reject 8.8\% of the time at the nominal 5\% level, and normal critical values scaled by the exact standard deviation reject 3.7\% (Fig.~\ref{fig:limit}(b)): the limit is more dispersed than the standard error suggests, but its tails are lighter than those of a normal distribution with the same variance.

As $K$ grows, the excess of the variance \eqref{eq:varTK} over one grows like $\rho^2(2K)^{1/2}/\pi$, so the ratio of the usual standard error to the true standard deviation decays like $K^{-1/4}$: it is 0.84, 0.79 and 0.73 for $K=2$, 5 and 10. The limit also becomes skewed to the left. When the training moments of all folds nearly sum to zero, $\sum_{r\neq k}Z_r\approx-Z_k$ for every $k$, so each fold is ranked against the sign of its own training moment and all $K$ terms of \eqref{eq:Trho} tend to be negative together. The two modes move apart as $K$ grows, to $-3.75$ and $0.70$ for $K=50$, and a symmetric 5\% rejection region then has almost all of its probability in the left tail (Table~\ref{tab:modes}).

The case $\rho=1$ arises in practice: after rescaling, it is the limit of the two-fold cross U-statistic of \citet{kim2024crossu}, and it is the limit of a test in common use. The best linear predictor test of \citet{chernozhukov2025generic} regresses the transformed outcome of a fold on its learned score. With a least-squares score on one covariate, the cross-fitted version takes each fold's own training moment, with the sign of the slope learned on the other folds, and divides it by its standard error. The evaluation moment of a fold is then its training moment, and the limit is $B_K=T_K(1)$. As the proof of Proposition~\ref{prop:shape}(ii) in the Appendix shows, $|B_2|$ is the larger of two independent half-normal variables, so the two-fold test with normal critical values rejects with probability $1-(1-\alpha)^2$, which is 9.75\% at $\alpha=0.05$; in our simulations it rejected 9.6\% of the time with two folds and 13.5\% with five, and joint calibration 4.9--5.0\% (\S~\ref{sec:numerics}).

\subsection{A published central limit theorem}\label{sec:published}

Theorem~6 of \citet{imai2025} asserts that the cross-fitted group effect, standardized by its true variance, is asymptotically normal. The model of \S~\ref{sec:univariate} satisfies each of its conditions as they are stated: Table~\ref{tab:assumptions} of the Supplementary Material sets them out one by one. In brief, the units are independent and identically distributed without interference, treatment is completely randomized with the same number of treated units in every fold, the potential outcomes are standard normal, and because no unit has an effect, the continuity condition on the effect along the score and the stability condition, Assumption~7, hold with every quantity identically zero. By Theorem~\ref{thm:counter} the standardized $\hat\theta_+$ is not asymptotically normal, so the model is a counterexample. Their Assumption~7 requires that, for every $a<b$,
\[
\bigl\|E\{Y(1)-Y(0)\mid a\leq s_{\mathcal T}(X)\leq b\}\bigr\|_{L^2(\mathcal T)}=o(m^{-1}),
\]
where $s_{\mathcal T}$ is the score learned on a training sample $\mathcal T$; here the left-hand side is identically zero. The step of their proof that fails, in their Supplementary Material~S9, uses this assumption to replace the group indicators of the learned ranking by those of a fixed ranking. The assumption controls the mean effect within groups, whereas the replacement needs $m^{1/2}P_k\{(\hat g_k-g^0)(\tilde Y-\tau)\}$, the difference between the evaluation moments under the learned weight and under a fixed weight $g^0$, to vanish. With $g^0(x)=\sign(x)$, the conditional second moment of this term given the training folds is $16$ when the learned slope is negative, which happens with probability one half (Lemma~\ref{lem:weight_l2}). No other fixed weight does better: the learned weight is $\sign(x)$ or $-\sign(x)$ with probability one half each, and a fixed weight differs from at least one of the two on a set of positive probability. Close scores need not give close rankings: $s$ and $cs$ rank units identically for $c>0$, but two slopes that both shrink to zero with opposite signs swap the groups.

The stability conditions of \citet{austern2020} and \citet{bayle2020cv}, which bound the change in the evaluated loss when the training sample changes, fail for the same reason. The central limit theorem of \citet{bayle2020cv} assumes that replacing one training point changes the centred loss by $o(\sigma_n^2/n)$ in mean square, where $\sigma_n^2$ is the variance of the loss averaged over training samples. Here that average is zero, so $\sigma_n^2=0$, while replacing one training point flips the sign of the slope with positive probability.

\subsection{Several covariates}\label{sec:directions}

With several covariates the learner chooses a direction instead of a sign, and the question is how much of the behaviour of \S~\ref{sec:univariate} survives. Let $X\sim N(0,I_d)$ in the model of that section, and let the ranking applied to fold $k$ be the least-squares score learned on the other folds. The training moment $Z_k$ is now a vector in $\mathbb R^d$, and the ranking depends on it only through the direction $b_k=S_k/|S_k|$ of the fitted coefficients, where $S_k=\sum_{r\neq k}Z_r$. With $Z_1,\ldots,Z_K$ independent $N(0,I_d)$ vectors and $W$ a standard normal variable independent of them, let
\[
B_{K,d}=K^{-1/2}\sum_{k=1}^KZ_k^\T b_k,\qquad T_{K,d}=\rho B_{K,d}+(1-\rho^2)^{1/2}W,\qquad \gamma_{K,d}=E\{(Z_1^\T b_1)(Z_2^\T b_2)\}.
\]
For $d=1$, $T_{K,1}$ has the distribution of $T_K(\rho)$.

\begin{proposition}\label{prop:dim}
In this model, with $K\geq2$ and $d\geq1$ fixed, $n^{1/2}D_n/4$ converges in distribution to $T_{K,d}$ with $\rho=(2/\pi)^{1/2}$, second moments converge, and $\var(T_{K,d})=1+\rho^2(K-1)\gamma_{K,d}$, where for $d\geq2$
\begin{equation}\label{eq:gamma}
\gamma_{K,d}=E\bigl[\{d-2+(b_1^\T b_2)^2\}/(|S_1|\,|S_2|)\bigr]>0.
\end{equation}
Moreover:
\begin{enumerate}
\item for every $K$, $d$ and $\rho\neq0$, no rescaling of $T_{K,d}$ is normal;
\item for $K=2$, $\gamma_{2,d}=a_d$, where $a_d=(E\chi_d)^2/d$ lies in $[d/(d+1),1)$ and $\chi_d$ is a chi random variable with $d$ degrees of freedom, and the fourth cumulant of $T_{2,d}$ is $3\rho^4\{d-1-2a_d-(d+2)a_d^2\}/(d+2)<0$;
\item for $d\geq3$ and every $K$, $(K-1)\gamma_{K,d}\leq(d-1)/(d-2)$, with equality in the limit as $K\to\infty$;
\item for fixed $K$, as $d\to\infty$, $T_{K,d}$ converges in distribution to $N(0,1+\rho^2)$ and $(K-1)\gamma_{K,d}\to1$.
\end{enumerate}
\end{proposition}

The proof in \S~\ref{app:dimproof} rests on the decomposition of Proposition~\ref{prop:structure}: the part of the evaluation moments that is uncorrelated with the training moments contributes an exactly normal variable with variance $1-\rho^2$, so all non-normality lies in $B_{K,d}$, and two Gaussian integrations by parts give \eqref{eq:gamma}.

\begin{figure}
\figuresize{.85}
\figurebox{20pc}{25pc}{}[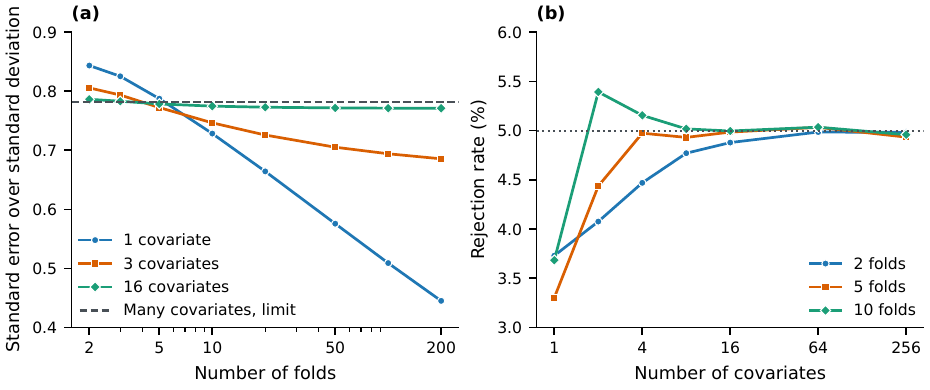]
\caption{With more covariates the limit under leave-one-fold-out approaches normality while its variance stays inflated; $\rho^2=2/\pi$. (a) Ratio of the usual standard error to the true standard deviation. (b) Rejection rate of a nominal 5\protect\% two-sided test with normal critical values scaled by the exact standard deviation.}
\par\smallskip{\footnotesize Alt text: Two line plots. Panel (a) shows the ratio of the usual standard error to the true standard deviation against the number of folds, from 2 to 200 on a log scale: with one covariate the ratio falls from 0.84 to 0.45, with three covariates from 0.81 to 0.69, and with sixteen covariates it stays near 0.78, the horizontal line marking the limit for many covariates. Panel (b) shows the rejection rate against the number of covariates, from 1 to 256 on a log scale, for 2, 5 and 10 folds: the rates start between 3.3\% and 3.7\% with one covariate, the 10-fold rate peaks at 5.4\% with two covariates, and all approach 5\% from about 16 covariates on.}
\label{fig:dimension}
\end{figure}

Parts~(iii) and~(iv) carry the practical message, and parts~(i) and~(ii) qualify it. By (iv), with many covariates the limit is close to normal but its variance is $1+\rho^2$, not one; by (iii), adding folds does not remove this excess once there are three or more covariates. By (i) and (ii), normality is nevertheless never exact, and with two folds the tails are lighter than normal for every $d$. Fig.~\ref{fig:dimension} shows the two regimes that result. With one covariate the learner has only a sign to choose, the limit is a mixture over signs, and its variance grows without bound as folds are added. With many covariates each training fold changes the direction only slightly and these changes average out: the variance tends to $1+\rho^2$ for every $K$, the ratio of the usual standard error to the true standard deviation is $(1+\rho^2)^{-1/2}=0.78$, and the limit is nearly normal. At $K=2$ the excess kurtosis falls from $-0.51$ with one covariate to $-0.10$ with sixteen, and normal critical values scaled by the exact standard deviation reject 3.7\% and 4.9\% of the time. Every limit we computed with two or more covariates is unimodal.

\section{Consequences for standard inference}\label{sec:consequences}

\subsection{Variance corrections}\label{sec:repairs}

Two repairs suggest themselves: inflating the sample variance of the fold estimates, and dividing by the exact standard deviation. Let $\hat\psi_k$ be the estimate from fold $k$ alone, $\hat\psi$ their mean and $S^2$ their sample variance. \citet{nadeau2003inference} replace $S^2/K$ by $(1/K+q)S^2$, where $q$ is the ratio of the sizes of the test and training sets, to account for overlapping training sets; under leave-one-fold-out $q=1/(K-1)$. If the $\hat\psi_k$ have a common variance and pairwise correlation $r$, then $E(S^2)=(1-r)\var(\hat\psi_1)$ and $\var(\hat\psi)=\{1+(K-1)r\}\var(\hat\psi_1)/K$, so
\begin{equation}\label{eq:nb}
\frac{E\{(1/K+q)S^2\}}{\var(\hat\psi)}=\frac{(1+Kq)(1-r)}{1+(K-1)r},
\end{equation}
which equals one only if $r=q/(1+q)=1/K$. With one covariate, $r=2\rho^2/\{\pi(2K-3)^{1/2}\}$ by Theorem~\ref{thm:counter}, so the correction overstates the variance when $4\rho^4K^2<\pi^2(2K-3)$ and understates it otherwise: it is conservative with few folds and anti-conservative with many. As $d\to\infty$, $r\to\rho^2/(K-1)$ by Proposition~\ref{prop:dim}, and the correction overstates the variance whenever $\rho^2<(K-1)/K$, which holds for every $K\geq3$ when $\rho^2=2/\pi$. The paired $t$-test, with $q=0$, always understates the variance when $r>0$. Dividing by the exact standard deviation restores the level as $d$ grows but not for $d=1$, where the rejection rate lies between 3.3\% at $K=5$ and 8.0\% at $K=50$ (Fig.~\ref{fig:dimension}(b) and Table~\ref{tab:references}). No variance correction repairs a non-normal limit.

With many covariates the limit is close to normal, and its variance can be estimated without refitting. By Proposition~\ref{prop:structure}, the evaluation moment of fold $k$ has conditional variance $1-\rho_k^2$ given the training moments, where $\rho_k^2$ is the squared multiple correlation between the evaluation moment of the selected ranking and the training moment of the fold, and by Proposition~\ref{prop:dim}(iv) the variance of the limit tends to $1+\rho^2$ as $d$ grows. We estimate $\rho_k^2$ by the squared multiple correlation between the contributions $\hat g_k(X_i)\hat\nu_i$ and $\{h(X_i)-\bar h\}\hat\nu_i$ of the units to the two moments, with $\hat\nu_i=\tilde Y_i-Q_n$ and covariances computed within arms, and divide $T_n$ by $(1+\bar\rho^{\,2})^{1/2}$, where $\bar\rho^{\,2}$ is the average of the estimates over folds. With one covariate this correction is conservative, since the excess variance $2\rho^2(K-1)/\{\pi(2K-3)^{1/2}\}$ is below $\rho^2$ for $K\leq5$ and the limit has light tails; \S~\ref{sec:trials} shows that it restores the level with all covariates of two experiments.

\subsection{Local alternatives}\label{sec:signal}

Away from the null hypothesis two things change: the learned ranking begins to follow the signal as well as the noise, and the quantity that the test estimates depends on which ranking was learned. To see both in the simplest case, keep two folds and one covariate, and let $Y(0)=\varepsilon-\theta_mX/2$ and $Y(1)=\varepsilon+\theta_mX/2$, so that $\tau(x)=\theta_mx$; let $\delta_m=m^{1/2}\theta_m/2$, the slope divided by the standard error of its estimate from one fold, and $c=E|X|=(2/\pi)^{1/2}$. With $s_k=\sign(\hat\beta_k)$, the group difference of the learned rankings is $\Delta^{\rm learn}_m=\theta_mc(s_1+s_2)$, and its average over training samples is $\Delta^{\rm alg}_m=2\theta_mc\,E(s_1)$. For finite $\delta$, let $s_k^\delta=\sign(Z_k+\delta)$, $\bar s=2\Phi(\delta)-1$,
\[
R_\delta=2^{-1/2}(s_2^\delta U_1+s_1^\delta U_2),\qquad A_\delta=R_\delta+2^{-1/2}\delta c\,(s_1^\delta+s_2^\delta-2\bar s),
\]
with $(Z_k,U_k)$ as in \eqref{eq:Trho} and $\rho=c$.

\begin{theorem}\label{thm:signals}
(i) If $\delta_m\to\delta\in\mathbb R$, then $n^{1/2}(D_n-\Delta^{\rm learn}_m)/4$ and $n^{1/2}(D_n-\Delta^{\rm alg}_m)/4$ converge jointly in distribution to $R_\delta$ and $A_\delta$, with convergence of second moments; for every finite $\delta$, neither limit is normal after standardization by its variance, and the two limits differ when $\delta\neq0$. (ii) If $\theta_m$ converges to a finite $\theta_*$ and $m^{1/2}|\theta_m|\to\infty$, then for both centrings $n^{1/2}(D_n-\Delta_m)/\{16+4\theta_*^2(1-c^2)\}^{1/2}$ converges in distribution to $N(0,1)$. (iii) If $\theta_m=2\delta/m^{1/2}$, the Hellinger affinity between the distributions of the data at $\delta_0$ and $\delta_1$ tends to $\exp\{-(\delta_1-\delta_0)^2/4\}>0$, so no estimator of $\delta$ is consistent.
\end{theorem}

In words, part~(i) concerns signals of the same order as the sampling error of the slope fitted on one fold, part~(ii) signals that dominate it, and part~(iii) says that two signals of the first kind give data sets that no procedure can tell apart with certainty, however large the sample. Non-normality is therefore not confined to the null hypothesis: it holds along every sequence of local alternatives, and a learner whose predictions converge need not give a ranking that converges. For such signals the estimate has two natural targets, which differ at first order. The first, $\Delta^{\rm learn}_m$, is the group difference of the two rankings actually applied, and is random because the rankings are; the second, $\Delta^{\rm alg}_m$, is its average over training samples, a property of the learning algorithm. Both vanish under the null hypothesis, so the distinction matters for confidence intervals rather than for tests. The normal approximation becomes valid once the signal dominates the sampling error of the slope, but by (iii) the data cannot show whether it does, so neither a large sample nor a consistent learner justifies a normal reference distribution. The bootstrap fails for the same reason at the boundary of a parameter space and for directionally differentiable functionals \citep{andrews2000inconsistency,fang2019inference}. Proofs and a figure of the two limiting variances are in \S~\ref{sec:signallimits}.

\section{Joint calibration}\label{sec:calibration}

Proposition~\ref{prop:structure} shows that the normal approximation should be applied to the training and evaluation moments before the ranking is selected, not to the fold estimates after it. Joint calibration does this by a parametric bootstrap of the moments, in three steps, and keeps the usual statistic. First, compute the moments $Z_{k,n}$ and $U_{k,n}$ of each fold and the statistic $T_n$ of \eqref{eq:Tn}, and estimate $\Sigma$ by the sample covariance $\hat\Sigma$ of the unit-level vectors $V_i$, with $\tau$ replaced by $Q_n$. Second, for $b=1,\ldots,N$, draw independent vectors $(Z_k^{*\T},U_k^{*\T})^\T\sim N(0,\hat\Sigma)$ $(k=1,\ldots,K)$ and compute $T_b^*$ from \eqref{eq:Tn} with the draws in place of the moments and $\hat\Sigma_{UU}$ unchanged, so that the draws pass through the same selection function and the same training sets as the data. Third, report the symmetric $p$-value $\hat p=\{1+\#(b:|T_b^*|\geq|T_n|)\}/(N+1)$, or the equal-tailed one; Algorithm~\ref{alg:calibration} of the Supplementary Material states the procedure in full and defines both $p$-values.

\begin{theorem}\label{thm:calibration}
Suppose that the conditions of Theorem~\ref{thm:general} hold and that $\hat\Sigma$ converges in probability to $\Sigma$. Let $\hat c_\alpha$ be the $(1-\alpha)$ quantile of $|T^*|$, where $T^*$ is computed as $T_\Sigma$ in \eqref{eq:TSigma} with $\hat\Sigma$ in place of $\Sigma$. If the distribution function of $|T_\Sigma|$ is continuous and strictly increasing at its $(1-\alpha)$ quantile, then $\pr(|T_n|>\hat c_\alpha)\to\alpha$ under the null hypothesis. The test that rejects when $T_n$ lies outside the $\alpha/2$ and $1-\alpha/2$ quantiles of $T^*$ has asymptotic level $\alpha$ when the distribution function of $T_\Sigma$ is continuous and strictly increasing at those quantiles.
\end{theorem}

The symmetric and equal-tailed regions coincide for two folds and an odd selection function, since replacing $(Z_1,U_1)$ by $(-Z_1,-U_1)$ maps $T_\Sigma$ to $-T_\Sigma$. With more folds $T_\Sigma$ is skewed to the left while a signal shifts $T_n$ to the right, so the symmetric region has most of its probability in the tail that a signal does not reach; with $K=50$ in the model of \S~\ref{sec:univariate}, a signal that the equal-tailed test detects 25.7\% of the time is detected 1.8\% of the time by the symmetric one. We therefore recommend the equal-tailed region whenever $K>2$. When the numbers of treated units in each fold are fixed, $\Sigma$ is replaced by a covariance computed within arms (\S~\ref{sec:design}). With heavy-tailed outcomes the sample covariance may be replaced by a robust estimator \citep{ledoit2004well,wei2017optimal,ke2019user,he2026cross}; Theorem~\ref{thm:calibration} requires only that the estimator be consistent. The test needs one pass through the data and no refitting: with $q=d+J$, its cost is of order $nq^2+q^3+NKq^2$, whereas a randomization test refits the learner for every draw of the treatment labels, and rank-transformed subsampling \citep{guo2025rts}, which applies to a wider class of procedures, refits it on every subsample.

\section{Numerical results}\label{sec:experiments}

\subsection{Simulations}\label{sec:numerics}

We simulated the model of \S~\ref{sec:univariate} with $n=6000$, splitting each fold at its empirical median. Table~\ref{tab:references} compares five reference distributions for the same statistic. Normal critical values with the usual standard error over-reject more as folds are added, in line with the rejection rates 8.8\%, 11.0\% and 15.1\% implied by the limits of Theorem~\ref{thm:counter} for $K=2$, 5 and 10. The correction of \citet{nadeau2003inference} holds the level up to $K=10$ and over-rejects from $K=20$, as \eqref{eq:nb} predicts, and where it holds the level it has little power. Normal critical values scaled by the exact standard deviation under-reject at small $K$ and over-reject at $K=50$. Joint calibration, with the covariance estimated in each replication, rejects 4.5--5.1\% at every $K$, and its equal-tailed version keeps its power as $K$ grows. The variance and kurtosis of the statistic are close to those of its limit, and the critical value of the limit holds the level with as few as 200 units (\S~\ref{app:gatessim}).

\begin{table}
\tbl{Rejection rates (\%) of two-sided tests of the group difference with one covariate and $n=6000$}{%
\begin{tabular}{rcccccccc}
& \multicolumn{5}{c}{Level} & \multicolumn{3}{c}{Power}\\
$K$ & Normal & Corrected & Exact & Joint, sym. & Joint, eq. & Corrected & Joint, sym. & Joint, eq.\\[2pt]
2 & 9.1 & 4.4 & 3.9 & 5.0 & 5.1 & 6.0 & 26.4 & 26.6\\
5 & 11.2 & 3.7 & 3.3 & 5.0 & 5.0 & 9.3 & 23.8 & 26.4\\
10 & 15.2 & 5.3 & 3.8 & 5.1 & 4.9 & 10.4 & 17.6 & 25.7\\
20 & 18.7 & 9.3 & 4.8 & 4.9 & 5.0 & 12.6 & 8.9 & 25.7\\
50 & 19.3 & 12.1 & 7.7 & 4.5 & 4.8 & 13.1 & 1.8 & 25.7
\end{tabular}}
\label{tab:references}
\begin{tabnote}
Corrected, variance correction of \citet{nadeau2003inference}; Exact, scaled by exact standard deviation; Joint, sym.\ and eq., joint calibration with 999 draws. Power is for $\tau(x)=4x/n^{1/2}$. Monte Carlo standard errors are at most 0.4 for levels and 0.8 for powers.
\end{tabnote}
\end{table}

Table~\ref{tab:refit} compares joint calibration, on the same data sets, with two reference distributions that refit the rankings: a randomization test, which redraws the treatment labels within folds 499 times, refits the rankings for every draw and is exact under the sharp null hypothesis of no effect on any unit, and rank-transformed subsampling \citep{guo2025rts}, which refits them on 800 subsamples of size $n/\log n$. All three held the level at every $K$, rejecting 4.3--5.6\% of the time. Their power depended on the rejection region more than on the method: with the symmetric region all three lost power as $K$ grew, for the reason given in \S~\ref{sec:calibration}, whereas with equal tails joint calibration and the randomization test kept about 25\% power at every $K$. Per data set, joint calibration took 2--4 milliseconds, the randomization test about 0.4 seconds and subsampling about 0.2 seconds. Each refit here is a closed-form slope; for a learner that is expensive to fit, the last two times grow with the cost of a fit and the first does not.

\begin{table}
\tbl{Rejection rates (\%) of two-sided tests with reference distributions that refit the rankings, one covariate and $n=6000$}{%
\begin{tabular}{rlccccc}
& & \multicolumn{2}{c}{Joint calibration} & \multicolumn{2}{c}{Randomization} & Subsampling\\
$K$ & & Sym. & Eq. & Sym. & Eq. & Sym.\\[2pt]
2 & Level & 4.6 & 4.8 & 4.4 & 4.3 & 4.9\\
 & Power & 27.2 & 27.4 & 27.2 & 26.3 & 28.1\\
5 & Level & 4.9 & 4.8 & 5.0 & 4.7 & 4.9\\
 & Power & 25.3 & 27.0 & 25.3 & 26.9 & 22.9\\
10 & Level & 4.9 & 5.0 & 4.7 & 4.7 & 5.0\\
 & Power & 16.7 & 25.0 & 15.9 & 23.8 & 13.7\\
20 & Level & 4.7 & 5.6 & 4.7 & 5.4 & 4.8\\
 & Power & 7.6 & 25.1 & 7.8 & 24.6 & 6.7
\end{tabular}}
\label{tab:refit}
\begin{tabnote}
Randomization, 499 label redraws within folds; Subsampling, rank-transformed subsampling \citep{guo2025rts} with 800 subsamples. Joint calibration uses 999 draws. Power is for $\tau(x)=4x/n^{1/2}$. Maximum Monte Carlo standard errors are 0.5 for levels and 1.4 for powers.
\end{tabnote}
\end{table}

Table~\ref{tab:settings} varies the setting, with two and five folds: binary outcomes; $t$-distributed errors with three degrees of freedom; 200 units; a fifth of the units treated; individual effects that vary while the conditional average effect is constant, so that the sharp null hypothesis fails; five covariates, of which the median split with the largest absolute training moment is selected, with the sign of that moment; and the best linear predictor test of \S~\ref{sec:univariate}. Normal critical values rejected 8.6--13.5\% of the time and the correction of \citet{nadeau2003inference} 3.5--5.2\%, while joint calibration rejected 3.9--5.9\% and the randomization test 4.1--6.2\%. With varying individual effects the randomization test is not exact, but it still held the level, as studentized permutation tests often do under weak null hypotheses \citep{chung2013exact}.

\begin{table}
\tbl{Rejection rates (\%) of two-sided 5\% tests under the null hypothesis in further settings}{%
\begin{tabular}{lrcccccc}
& & & & \multicolumn{2}{c}{Joint calibration} & \multicolumn{2}{c}{Randomization}\\
Setting & $K$ & Normal & Corrected & Sym. & Eq. & Sym. & Eq.\\[2pt]
Binary outcome & 2 & 9.3 & 4.7 & 4.9 & 4.7 & 4.7 & 4.5\\
 & 5 & 10.8 & 3.6 & 5.1 & 4.9 & 4.5 & 4.2\\
Heavy tails & 2 & 9.5 & 3.7 & 5.7 & 5.4 & 5.2 & 5.0\\
 & 5 & 9.8 & 3.5 & 4.1 & 3.9 & 4.5 & 4.2\\
$n=200$ & 2 & 9.4 & 4.7 & 5.5 & 5.5 & 5.0 & 5.0\\
 & 5 & 12.0 & 3.6 & 5.5 & 5.1 & 5.2 & 4.9\\
$p=0.2$ & 2 & 9.4 & 4.2 & 5.7 & 5.9 & 6.2 & 5.9\\
 & 5 & 11.0 & 3.7 & 5.1 & 5.1 & 5.3 & 5.1\\
Varying effects & 2 & 9.0 & 5.2 & 4.9 & 5.0 & 5.1 & 5.0\\
 & 5 & 10.5 & 3.5 & 5.1 & 4.8 & 4.8 & 4.3\\
Five rankings & 2 & 8.6 & 3.5 & 4.9 & 4.5 & 5.1 & 4.8\\
 & 5 & 11.3 & 3.7 & 4.6 & 4.7 & 4.5 & 4.1\\
Best linear predictor & 2 & 9.6 & -- & 4.9 & 5.0 & -- & --\\
 & 5 & 13.5 & -- & 5.0 & 5.0 & -- & --
\end{tabular}}
\label{tab:settings}
\begin{tabnote}
Corrected, \citet{nadeau2003inference}; Randomization, 499 redraws within folds. Baseline: $n=2000$, one covariate, $p=0.5$. Heavy tails, $t_3$ errors; Varying effects, individual effects with variance 4; Five rankings, 5 covariates; Best linear predictor, $n=6000$. Monte Carlo standard errors are at most 0.5 near 5\% and 0.8 otherwise.
\end{tabnote}
\end{table}

\subsection{An e-mail experiment}\label{sec:real}

The Hillstrom experiment randomized customers to a men's e-mail, a women's e-mail or none \citep{hillstrom2008,hillstromdata}. We compare the men's e-mail with no e-mail for 42\,613 customers, with a website visit within two weeks as outcome, and rank customers by the logarithm of one plus their spending in the past year: each of two folds learns a least-squares slope and ranks the other. The standardized statistic is 2.11. Treating the folds as independent gives a two-sided $p$-value of 0.035, whereas joint calibration gives 0.065, and rank-transformed subsampling, at a much higher computational cost, gives 0.054--0.064. The reference distributions disagree because the training and evaluation moments of each fold have correlation about 0.85. Two fold statistics with such correlations $\rho_1$ and $\rho_2$ have correlation $2\rho_1\rho_2/\pi$ in the limit, here 0.46, and the correlation observed across subsamples is the same. With five and ten folds the normal $p$-values are 0.047 and 0.029, and the equal-tailed joint $p$-values 0.092 and 0.057. To check which reference distribution is appropriate for these data, we redrew the e-mail labels 2000 times, so that the labels have no effect on any customer, and repeated the analysis: normal critical values rejected 8.8\%, 11.0\% and 16.8\% of the time with two, five and ten folds, and joint calibration 4.5--5.0\% (\S~\ref{app:hillstrom}).

\subsection{Experiments with redrawn treatment labels}\label{sec:trials}
To see the two regimes of Proposition~\ref{prop:dim} on real covariates, we kept the covariates and outcomes of two experiments and redrew the treatment labels, so that no unit has an effect and every rejection is false. In the ACTG~175 trial \citep{actgdata,hammer1996}, 1054 patients received zidovudine or zidovudine plus didanosine; the outcome is the change in CD4 count and there are 16 baseline covariates. From the Criteo uplift data \citep{criteodata,diemert2018} we took 20\,000 rows with 12 features and a website visit as outcome. With a least-squares ranking on a single covariate the statistic had clearly negative excess kurtosis, and the test standardized by its exact standard deviation was conservative; with all covariates the statistic was close to normal, as Proposition~\ref{prop:dim} predicts, and the same test had the nominal level. When the ranking is selected from a finite set of median splits, by the largest absolute training moment and with its sign, Theorem~\ref{thm:calibration} applies: joint calibration rejected 4.5--5.9\% of the time, where normal critical values rejected 8.6--13.2\%. For least squares of the transformed outcome on all covariates, the correction of \S~\ref{sec:repairs} rejected 3.2--5.2\% of the time, where normal critical values rejected 10.7--13.4\%; with a single covariate it was conservative, as expected (\S~\ref{app:realtrials}).

\section{Discussion}\label{sec:discussion}

Four questions remain open. First, joint calibration draws a vector of moments of fixed, finite dimension. Rankings learned from infinite classes, such as least squares on many covariates or a random forest, at present need a reference distribution that refits the learner, as rank-transformed subsampling and the randomization test do; a calibration of such rankings without refitting would be useful. The correction of \S~\ref{sec:repairs} is one for least squares on many covariates, but its justification rests on the Gaussian model of Proposition~\ref{prop:dim} and has yet to be extended to other covariate distributions. Second, under local alternatives the group effect of the learned ranking and its average over training samples differ at first order, and uniform inference for the latter cannot rely on an estimate of the signal, which by Theorem~\ref{thm:signals}(iii) cannot be estimated consistently. Third, our limits hold the numbers of candidate rankings, covariates and folds fixed as the sample grows. The order of the limits matters: with one covariate, $T_K(\rho)$ converges in distribution to a skew-normal distribution as $K\to\infty$, but its variance diverges like $K^{1/2}$, so its moments do not converge, and the rejection rate of the test standardized by the exact standard deviation is not monotone in $K$ (Table~\ref{tab:references}). With one unit per fold, $K=n$, there is neither a central limit theorem nor a median within a fold, and neither Theorem~\ref{thm:general} nor joint calibration applies. Fourth, the statistic is often averaged over repeated partitions. Methods that combine split-sample statistics \citep{meinshausen2009pvalues,diciccio2020exact,ritzwollerromano} take as inputs single-partition statistics with known reference distributions, which joint calibration supplies, but the distribution of the average over partitions has not been described.

\section*{Appendix: Proof of Proposition~\ref{prop:shape}}

(ii) Because $B_2$ is symmetric we may take $\rho\geq0$. By (i) and $\sign(Z_2)Z_1+\sign(Z_1)Z_2=\sign(Z_1Z_2)(|Z_1|+|Z_2|)$, $B_2=\epsilon R$ with $\epsilon=\sign(Z_1Z_2)$ independent of $R=(|Z_1|+|Z_2|)/2^{1/2}$. Hence the characteristic function of $B_2$ is $E\cos(uR)$, the real part of $\chi(u/2^{1/2})^2$, where $\chi(s)=E\exp(is|Z_1|)=\exp(-s^2/2)+i(2/\pi^{1/2})\daw(s/2^{1/2})$; this gives $\exp(-u^2/2)-(4/\pi)\daw(u/2)^2$ and, with (i), \eqref{eq:cf2}. The fourth cumulant follows from $E(R^2)=1+2/\pi$, $E(R^4)=3+8/\pi$ and the scaling of cumulants. For the density, $|x|+|y|=\max(|x+y|,|x-y|)$ shows that $R=\max(|W_1|,|W_2|)$ with $W_1=(Z_1+Z_2)/2^{1/2}$ and $W_2=(Z_1-Z_2)/2^{1/2}$ independent standard normal, so $B_2$ has density $4\phi(y)\Phi(|y|)-2\phi(y)$. The density of $T_2(\rho)$ is the convolution of the law of $\rho B_2$ with $N(0,\sigma^2)$. The term $-2\phi$ contributes $-2\phi(x)$, because $\rho^2+\sigma^2=1$. For the other term, the $N(0,\rho^2)$ density times the $N(x,\sigma^2)$ density, both at $y$, equals $\phi(x)$ times the $N(\rho^2x,\rho^2\sigma^2)$ density at $y$, so the part with $y>0$ contributes $4\phi(x)\pr(Y>0,\ Y>\rho G)$, where $Y=\rho^2x+\rho\sigma V$ and $V$ and $G$ are independent standard normal variables. Now $Y>0$ if and only if $-V<ax$, and $Y>\rho G$ if and only if $(G-\sigma V)/(1+\sigma^2)^{1/2}<bx$; the two left-hand sides are standard normal with correlation $r$, so the probability is $\Phi_r(ax,bx)$. The part $y<0$ gives $4\phi(x)\Phi_r(-ax,-bx)$, proving \eqref{eq:density2}.

(iii) Write $f_\rho=2\phi g$, with $g(x)=2\Phi_r(ax,bx)+2\Phi_r(-ax,-bx)-1$ even, so that $f_\rho''(0)=2\phi(0)\{g''(0)-g(0)\}$. Since $\partial\Phi_r(u,v)/\partial u=\phi(u)\Phi\{(v-ru)/(1-r^2)^{1/2}\}$, and symmetrically in $v$, the second derivative of $x\mapsto\Phi_r(ax,bx)$ at zero is $\phi(0)^2\{2ab-r(a^2+b^2)\}/(1-r^2)^{1/2}=\phi(0)^2\rho^2/(\sigma+\sigma^3)$, and the same holds for $\Phi_r(-ax,-bx)$. Also $\Phi_r(0,0)=1/4+(\arcsin r)/(2\pi)$ and $\arcsin r=\arctan\sigma$, so $g(0)=(2/\pi)\arctan\sigma$ and $g''(0)=(2/\pi)\rho^2/(\sigma+\sigma^3)$, which gives the expression for $f_\rho''(0)$. As a function of $\sigma\in(0,1)$, the bracket is $(1-\sigma^2)/(\sigma+\sigma^3)-\arctan\sigma$; its first term decreases and its second increases, from $\infty$ as $\sigma\to0$ to $-\pi/4$ at $\sigma=1$, with unique zero $\sigma_*=0.6635$, so $\rho_*=0.7482$.

\bibliographystyle{plainnat}
\bibliography{references}

\clearpage
\appendix

\begin{center}
{\LARGE\bfseries Supplementary Material for\\[0.35em]
``Non-normal Limits and Joint Calibration of Cross-fitted Heterogeneity Tests''}
\end{center}
\vspace{1.5em}

\setcounter{section}{0}
\renewcommand{\thesection}{S\arabic{section}}
\renewcommand{\theHsection}{S\arabic{section}}

\setcounter{table}{0}
\renewcommand{\thetable}{S\arabic{table}}
\renewcommand{\theHtable}{S\arabic{table}}

\setcounter{figure}{0}
\renewcommand{\thefigure}{S\arabic{figure}}
\renewcommand{\theHfigure}{S\arabic{figure}}

\setcounter{equation}{0}
\renewcommand{\theequation}{S\arabic{equation}}
\renewcommand{\theHequation}{S\arabic{equation}}

\setcounter{lemma}{0}
\renewcommand{\thelemma}{S\arabic{lemma}}
\renewcommand{\theHlemma}{S\arabic{lemma}}

\setcounter{algocf}{0}
\renewcommand{\thealgocf}{S\arabic{algocf}}
\renewcommand{\theHalgocf}{S\arabic{algocf}}

Equations, theorems, tables and figures numbered without the prefix S refer to the paper. Section~\ref{app:notation} restates the definitions that the proofs cite; \S\S~\ref{app:limits}--\ref{app:calibration} prove the results of \S\S~\ref{sec:general}--\ref{sec:calibration}; \S~\ref{app:experiments} gives the designs and complete tables of \S~\ref{sec:experiments}.

\section{Notation and models}\label{app:notation}

This section restates, with equation numbers, the definitions and models of the paper that the proofs below cite. Throughout, $P_kv=m^{-1}\sum_{i\in I_k}v_i$ and $P_nv=n^{-1}\sum_{i=1}^nv_i$. The transformed outcome and the conditional average treatment effect are
\begin{equation}\label{eq:tildeY}
\tilde Y=AY/p-(1-A)Y/(1-p),\qquad \tau(x)=E(\tilde Y\mid X=x).
\end{equation}
The null hypothesis of no heterogeneity, $\tau(X)=\tau$ almost surely for a constant $\tau$, is written $H_\tau$.
A score $s_{\calT}$ learned on a training sample $\calT$ splits units at its median into halves with indicators $w_{\calT,+}$ and $w_{\calT,-}$, and the group effects of the learned ranking are
\begin{equation}\label{eq:target}
\theta_\pm(\calT)=2E[w_{\calT,\pm}(X)\{Y(1)-Y(0)\}\mid\calT].
\end{equation}
In fold $k$ the empirical median gives weights $\hat w_{k,\pm}$ and the centred weight $\hat g_k=2\hat w_{k,+}-1$; then $\hat\theta_{k,\pm}=2P_k(\hat w_{k,\pm}\tilde Y)$, $D_k=2P_k(\hat g_k\tilde Y)$, $\hat\theta_\pm$ and $D_n$ are the averages over folds, and with $Q_n=P_n\tilde Y$,
\begin{equation}\label{eq:bridge}
\hat\theta_\pm=Q_n\pm D_n/2.
\end{equation}
The usual standard error treats the contributions $\Gamma_i=2\hat g_k(X_i)\tilde Y_i$ $(i\in I_k)$ as independent:
\begin{equation}\label{eq:est}
s_n^2=n^{-1}\Bigl\{n^{-1}\sum_{i=1}^n\Gamma_i^2-D_n^2\Bigr\}.
\end{equation}

The model of \S~\ref{sec:univariate} has $X_i$ and $\varepsilon_i$ independent standard normal and
\begin{equation}\label{eq:dgp}
Y_i(1)=Y_i(0)=\varepsilon_i,
\end{equation}
half of the units treated by complete randomization, and least squares without intercept in each fold,
\begin{equation}\label{eq:ols}
\hat\beta_k=\sum_{i\in I_k}X_i\tilde Y_i\Big/\sum_{i\in I_k}X_i^2,\qquad s_{I_k}(x)=\hat\beta_kx.
\end{equation}
In this model $r_k$ denotes the weight that equals $1$ for the units of fold $k$ whose covariate lies above the empirical median of the fold and $-1$ for the others, so that $\hat g_k=\sign(\hat\beta)\,r_k$ when $\hat\beta$ is the slope learned for fold $k$. The raw moments of the model are
\begin{equation}\label{eq:primitive_counter}
(Q_{k,m},Z_{k,m},U_{k,m})=(m^{1/2}/2)\,P_k\{\tilde Y,\ X\tilde Y,\ \sign(X)\tilde Y\}.
\end{equation}
Write $T_K=T_K\{(2/\pi)^{1/2}\}$ and $\widetilde T_K=T_K(\pi^{-1/2})$, with $T_K(\rho)$ as in \eqref{eq:Trho}. For two folds, Proposition~\ref{prop:shape} gives
\begin{equation}\label{eq:mixture}
T_2(\rho)=\rho B_2+(1-\rho^2)^{1/2}W,\qquad B_2=\sign(Z_1Z_2)(|Z_1|+|Z_2|)/2^{1/2},
\end{equation}
in distribution, with $W$ standard normal and independent of $B_2$, and
\begin{equation}\label{eq:variance}
\var\{T_2(\rho)\}=1+2\rho^2/\pi,
\end{equation}
\begin{equation}\label{eq:k4}
\kappa_4\{T_2(\rho)\}=-(4/\pi+12/\pi^2)\rho^4.
\end{equation}
The last statement of Theorem~\ref{thm:counter} reads
\begin{equation}\label{eq:truevar_limit}
\hat\theta_+/\var(\hat\theta_+)^{1/2}\to\widetilde T_K/\var(\widetilde T_K)^{1/2}\neq N(0,1)
\end{equation}
in distribution. Assumption~7 of \citet{imai2025} requires, for every $a<b$,
\begin{equation}\label{eq:ILcondition}
\bigl\|E\{Y(1)-Y(0)\mid a\leq s_{\calT}(X)\leq b\}\bigr\|_{L^2(\calT)}=o(m^{-1}).
\end{equation}

The general setting of \S~\ref{sec:generallimit} uses, with $\nu=\tilde Y-\tau$,
\begin{equation}\label{eq:V}
V=\bigl[\{h(X)-E h(X)\}^\T\nu,\ g_1(X)\nu,\ \ldots,\ g_J(X)\nu\bigr]^\T,
\end{equation}
\begin{equation}\label{eq:moments_general}
E[\{1+\|h(X)\|^2\}(1+\nu^2)]<\infty,\qquad \Sigma=E(VV^\T),
\end{equation}
\begin{equation}\label{eq:ZUhat}
Z_{k,n}=m^{1/2}P_k\{(h-P_kh)\tilde Y\},\qquad U_{k,n,j}=m^{1/2}P_k(\hat g_{kj}\tilde Y)\quad(j=1,\ldots,J),
\end{equation}
and the conditions of \S~\ref{sec:quantile} give
\begin{equation}\label{eq:expansions}
(Z_{k,n},U_{k,n})=m^{-1/2}\sum_{i\in I_k}V_i+o_p(1).
\end{equation}

The local alternatives of \S~\ref{sec:signal} keep two folds and one covariate, with
\begin{equation}\label{eq:signal_dgp}
Y(0)=\varepsilon-\theta_mX/2,\qquad Y(1)=\varepsilon+\theta_mX/2,
\end{equation}
so that $\tau_m(x)=\theta_mx$, $\delta_m=m^{1/2}\theta_m/2$ and $c=E|X|$. The fold coefficients satisfy
\begin{equation}\label{eq:signal_t}
m^{1/2}(\hat\beta_k-\theta_m)/2\sim t_m,\qquad \bar s_m=E\{\sign(\hat\beta_k)\}=2F_{t_m}(\delta_m)-1,
\end{equation}
where $t_m$ and $F_{t_m}$ denote Student's $t$ distribution with $m$ degrees of freedom and its distribution function. With $s_k=\sign(\hat\beta_k)$, the group differences of the learned rankings and of their average over training samples are
\begin{equation}\label{eq:signal_targets}
\Delta^{\rm learn}_m=\theta_mc(s_1+s_2),\qquad \Delta^{\rm alg}_m=2\theta_mc\,\bar s_m.
\end{equation}
For finite $\delta$, with $s_k^\delta=\sign(Z_k+\delta)$ and $\bar s=2\Phi(\delta)-1$,
\begin{equation}\label{eq:Rdelta}
R_\delta=2^{-1/2}(s_2^\delta U_1+s_1^\delta U_2),\qquad A_\delta=R_\delta+2^{-1/2}\delta c\,(s_1^\delta+s_2^\delta-2\bar s).
\end{equation}
With $\gamma=2\rho\phi(\delta)$, their variances are $\var(R_\delta)=1+\gamma^2$ and
\begin{equation}\label{eq:varlocala}
\var(A_\delta)=1+\gamma^2+\delta^2c^2(1-\bar s^2)+2\delta c\bar s\gamma.
\end{equation}
With $\theta_m=2\delta/m^{1/2}$, the Hellinger affinity of the distributions of the data at two drifts is
\begin{equation}\label{eq:affinity}
\mathcal A(P_{\delta_0,m},P_{\delta_1,m})=\bigl\{1+(\delta_1-\delta_0)^2/(4m)\bigr\}^{-m}\to\exp\{-(\delta_1-\delta_0)^2/4\}.
\end{equation}

\section{Asymptotic distributions}\label{app:limits}

\subsection{Conditional centring}\label{app:center}

Let $\cG_k$ be the $\sigma$-field generated by the fold labels, any randomness of the learner, all data outside fold $k$ and the covariates of fold $k$.

\begin{lemma}[Conditional centring]\label{lem:center}
Suppose that the ranking weight $\widehat g_k$ that evaluates fold $k$ is determined by the treatments and outcomes of the folds in $F(k)$, the covariates of all folds, the fold labels and the learner's randomness. Then, for $i\in I_k$, $\E[\widehat g_k(X_i)\tilde Y_i\mid\cG_k]=\widehat g_k(X_i)\tau(X_i)$, and these products are conditionally independent across the units of fold $k$. Under $H_\tau$, if the weights sum to zero over the fold, $D_k=2P_k(\widehat g_k\tilde Y)$ has conditional mean zero.
\end{lemma}

\begin{proof}
The weight $\widehat g_k$ is $\cG_k$-measurable: it uses treatments and outcomes only from the folds in $F(k)$, and covariates, including those of fold $k$ through the empirical median. Given $\cG_k$, the pairs $(A_i,Y_i)$, $i\in I_k$, are independent, and randomization gives $\E(\tilde Y_i\mid\cG_k)=\E(\tilde Y_i\mid X_i)=\tau(X_i)$. Under $H_\tau$ the conditional mean of $D_k$ is $2\tau P_k\widehat g_k$, which is zero when the weights sum to zero.
\end{proof}

\subsection{Proof of Theorem~\ref{thm:counter} for two folds}\label{app:counterproof}

\textit{Step 1: the assignment and the moments.}
Condition on the assignment and on the split of each arm into folds. The assignment is independent of the potential outcomes and covariates, and the units are independent and identically distributed, so the units can be relabelled by fold and arm, and the pairs $(X_i,\varepsilon_i)$ remain independent with the distribution~\eqref{eq:dgp}. Since $\varepsilon_i$ is symmetric about zero, the variables $\xi_i=(2A_i-1)\varepsilon_i$ are, given the assignment, independent standard normal and independent of the $X_i$, and $\tilde Y_i=2\xi_i$.

The moments~\eqref{eq:primitive_counter} are standardized sums of the independent vectors
\[
 \frac1{{m}^{1/2}}\sum_{i\in I_k}
 (\xi_i,\ X_i\xi_i,\ \sgn(X_i)\xi_i)\trans,
\]
with mean zero and covariance
\begin{equation}\label{eq:countercov}
 \begin{pmatrix}
 1&0&0\\
 0&1&(2/\pi)^{1/2}\\
 0&(2/\pi)^{1/2}&1
 \end{pmatrix}.
\end{equation}
The multivariate central limit theorem gives joint convergence within each fold, and the two folds are independent. The conditional distribution given the assignment does not depend on which units are treated, so the same limit holds unconditionally. For the same reason, simulations in which the treated units are fixed have the distribution of the statistic under complete randomization.

\textit{Step 2: replacing the empirical median.}
Let $N_k^+=\sum_{i\in I_k}\ind\{X_i>0\}$. Because $X$ is continuous, the empirical upper half of fold $k$ and the population upper half differ in exactly
\[
 C_k=|N_k^+-m/2|=\Op({m}^{1/2})
\]
units. Write
\[
 R_k=\frac1{2{m}^{1/2}}\sum_{i\in I_k}
 \{r_k(X_i)-\sgn(X_i)\}\tilde Y_i.
\]
Given the covariates of the fold, $R_k$ has mean zero and only $C_k$ non-zero coefficients, each of absolute value 2. Hence
\begin{equation}\label{eq:medianbound}
 \E(R_k^2\mid X_{I_k})=\frac{4C_k}{m},\qquad
 \E R_k^2\le\frac{2}{{m}^{1/2}}\longrightarrow0,
\end{equation}
where the last inequality uses $N_k^+\sim\mathrm{Binomial}(m,1/2)$ and the Cauchy--Schwarz inequality. Multiplying by the sign learned in the other fold does not change the absolute value of the remainder, and the sum of the remainders over finitely many folds is $\op(1)$; no independence between the remainders of the two folds is used.

\textit{Step 3: the limit of the group difference.}
The denominator of~\eqref{eq:ols} is strictly positive, so $\sgn(\widehat\beta_k)=\sgn(Z_{k,m})$. The limit $Z_k$ is standard normal, so it equals zero with probability zero and the sign function is almost surely continuous at it. The joint central limit theorem,~\eqref{eq:medianbound} and the continuous mapping theorem give
\begin{align*}
 \frac{{n}^{1/2}D_n}{4}
 &=\frac1{{2}^{1/2}}\sum_{k=1}^2
 \sgn(Z_{3-k,m})\{U_{k,m}+R_k\}\\
 &\dto\frac{\sgn(Z_2)U_1+\sgn(Z_1)U_2}{{2}^{1/2}}=T_2.
\end{align*}

\textit{Step 4: the effect in the upper half.}
Let $Q^*=(Q_1+Q_2)/{2}^{1/2}$. By~\eqref{eq:countercov}, $Q^*\sim\N(0,1)$ is independent of $(Z_1,U_1,Z_2,U_2)$, and
\[
 \frac{{n}^{1/2}Q_n}{2}=\frac{Q_{1,m}+Q_{2,m}}{{2}^{1/2}}\dto Q^*.
\]
Together with~\eqref{eq:bridge},
\[
 \frac{{n}^{1/2}\widehat\theta_+}{2{2}^{1/2}}
 \dto\frac{Q^*+T_2}{{2}^{1/2}}.
\]
By Proposition~\ref{prop:shape}(i), the right side equals
\[
 \frac{\rho}{{2}^{1/2}}B_2
 +\frac{Q^*+(1-\rho^2)^{1/2}W}{{2}^{1/2}}
 \overset d=\frac{\rho}{{2}^{1/2}}B_2+(1-\rho^2/2)^{1/2}\,W^*,
\]
with $W^*$ standard normal and independent of $B_2$, which is $\widetilde T_2$. The lower group has the same marginal limit by symmetry.

\textit{Step 5: convergence of the exact variance.}
Given the slope learned on the other fold and the covariates of fold $k$, the estimated effect in the upper half of fold $k$ is a linear combination of the $m/2$ independent standard normal variables $\xi_i$ of that half, and is exactly $\N(0,8/m)$. Hence for any fixed $r>4$,
\[
 \sup_m\E\left|\frac{{n}^{1/2}\widehat\theta_+}{2{2}^{1/2}}\right|^r<\infty,
\]
by Minkowski's inequality applied to the two single-fold estimates; their dependence is not used. The squares are therefore uniformly integrable, and weak convergence implies convergence of second moments. Since $\E\widehat\theta_+=0$ in finite samples,
\[
 \frac n8\Var(\widehat\theta_+)\longrightarrow1+\rho^2/\pi.
\]
Likewise $\frac n{16}\Var(D_n)\to1+2\rho^2/\pi$. Slutsky's theorem and~\eqref{eq:k4} give the non-normal limit~\eqref{eq:truevar_limit}.

\textit{Step 6: the estimands and the conditions of the published theorem.}
Because $Y_i(1)-Y_i(0)=0$ for every unit, the group effects~\eqref{eq:target} and their averages over training samples are zero, and so is the mean effect within every interval of the score. Given $X$, the least-squares numerator has a non-degenerate continuous normal distribution, so the coefficient is non-zero almost surely. Moreover
\[
 \E(\widehat\beta_k^2)=4\E\left[\left(\sum_{i\in I_k}X_i^2\right)^{-1}\right]
 =\frac4{m-2}\longrightarrow0,
\]
so the score is consistent in mean square and, its slope being non-zero, is one-to-one in every sample. Together with the remaining conditions in Table~\ref{tab:assumptions}, this completes the proof.

\subsection{Moments and tail probabilities of the two-fold limit}\label{app:moments}

Write $U_k=\rho Z_k+(1-\rho^2)^{1/2}E_k$ with $Z_1,Z_2,E_1,E_2$ independent standard normal. For $\rho\in[-1,1]$,
\begin{align*}
 T_2(\rho)
 &=\rho\frac{\sgn(Z_2)Z_1+\sgn(Z_1)Z_2}{{2}^{1/2}}
   +(1-\rho^2)^{1/2}\frac{\sgn(Z_2)E_1+\sgn(Z_1)E_2}{{2}^{1/2}}\\
 &=\rho B_2+(1-\rho^2)^{1/2}W.
\end{align*}
Given $Z_1,Z_2$, the variable $W$ is standard normal with a conditional distribution that does not depend on $Z_1,Z_2$, so $W$ is independent of $B_2$. The sign $\sgn(Z_1Z_2)$ is independent of the two absolute values and equals $\pm1$ with probability $1/2$ each, so $\E B_2=0$.

The absolute moments $\E|Z|=(2/\pi)^{1/2}$, $\E|Z|^2=1$, $\E|Z|^3=2(2/\pi)^{1/2}$ and $\E|Z|^4=3$ give
\[
 \E B_2^2=1+2/\pi,\qquad
 \E B_2^4=3+8/\pi,
\]
and therefore
\[
 \kappa_4(B_2)=3+8/\pi-3(1+2/\pi)^2
 =-4/\pi-12/\pi^2.
\]
Cumulants of independent variables add and the fourth cumulant of a normal variable is zero, which gives~\eqref{eq:variance}--\eqref{eq:k4}. For $\rho\neq0$ the fourth cumulant is non-zero, which rules out every normal distribution; for $\rho=0$, $T_2(0)=W$. Rescaling multiplies the fourth cumulant by the fourth power of the scale and so cannot make it zero.

The same representation gives the correlation between the two fold statistics that \S~\ref{sec:real} uses. With $U_k=\rho_kZ_k+(1-\rho_k^2)^{1/2}E_k$, independence of the folds gives $\E\{\sgn(Z_2)U_1\sgn(Z_1)U_2\}=\rho_1\rho_2\{\E|Z|\}^2=(2/\pi)\rho_1\rho_2$, and each product $\sgn(Z_{3-k})U_k$ has unit variance.

\textit{Tail probabilities.} With $R=|B_2|$,
\begin{equation}\label{eq:Rlaw}
 \Pp(R\le r)=\{2\Phi(r)-1\}^2,\qquad f_R(r)=4\phi(r)\{2\Phi(r)-1\},\quad r\ge0,
\end{equation}
so that for $0\le\rho<1$,
\begin{equation}\label{eq:tail}
 \Pp(|T_2(\rho)|>c)=\int_0^\infty f_R(r)\left[\overline\Phi\!\left(\frac{c-\rho r}{(1-\rho^2)^{1/2}}\right)+\overline\Phi\!\left(\frac{c+\rho r}{(1-\rho^2)^{1/2}}\right)\right]dr.
\end{equation}
To see this, let $V=(Z_1+Z_2)/{2}^{1/2}$ and $W_0=(Z_1-Z_2)/{2}^{1/2}$, which are independent standard normal, with
\[
 |B_2|=\frac{|Z_1|+|Z_2|}{{2}^{1/2}}=\max(|V|,|W_0|).
\]
Its distribution function is therefore the product of two half-normal distribution functions. Conditioning~\eqref{eq:mixture} on $|B_2|=r$, using the symmetry of its sign, and integrating gives~\eqref{eq:tail}. For $\rho=1$, $\Pp\{|T_2(1)|>c\}=1-\{2\Phi(c)-1\}^2$.

Table~\ref{tab:analytic} evaluates~\eqref{eq:tail} at four correlations: $\rho=0$ is the normal case, $\rho=(2/\pi)^{1/2}$ arises from a normal covariate, $\rho={3}^{1/2}/2$ from a covariate uniform on $[-1,1]$, and $\rho=1$ is the case in which the two moments coincide. For the effect in the upper half with a normal covariate, where $\rho=\pi^{-1/2}$, normal critical values scaled by the exact standard deviation reject with probability $4.70\%$.

\begin{table}[htbp]
\centering\small
\caption{Variance, kurtosis, quantile and rejection rates (\protect\%) of nominal 5\protect\% two-sided tests for the two-fold limit $T_2(\rho)$}\label{tab:analytic}
\setlength{\tabcolsep}{5pt}
\begin{tabular}{lrrrrr}
\toprule
$\rho$ & Variance & Kurtosis & 95\% quantile of $|T_2(\rho)|$ & Normal & Normal scaled by exact SD\\
\midrule
$0$ & 1.000 & 3.000 & 1.960 & 5.00 & 5.00 \\
$(2/\pi)^{1/2}$ & 1.405 & 2.489 & 2.206 & 8.82 & 3.73 \\
${3}^{1/2}/2$ & 1.477 & 2.359 & 2.221 & 9.21 & 3.30 \\
$1$ & 1.637 & 2.071 & 2.236 & 9.75 & 2.42 \\
\bottomrule
\end{tabular}
\end{table}

\subsection{Proof of Theorem~\ref{thm:counter} for \texorpdfstring{$K$}{K} folds}\label{app:kfoldproof}

Let the $K$ folds have equal size $m$ with $m/2$ treated units each. Steps~1--2 of \S~\ref{app:counterproof} hold fold by fold: the raw moments $(Q_{k,m},Z_{k,m},U_{k,m})$ of~\eqref{eq:primitive_counter} converge jointly to independent copies of $(Q_k,Z_k,U_k)$ with covariance~\eqref{eq:countercov}, and replacing the population median by the empirical median changes each $U_{k,m}$ by $\op(1)$, by~\eqref{eq:medianbound}. The slope learned on the other folds has the sign of $\sum_{r\ne k}Z_{r,m}$, because its denominator is positive. Hence
\[
\frac{{n}^{1/2}D_n}4=\frac{{Km}^{1/2}}{4K}\sum_{k=1}^K2\,\sgn\Big(\sum_{r\ne k}Z_{r,m}\Big)P_k(r_k\tilde Y)
=\frac1{{K}^{1/2}}\sum_k\sgn\Big(\sum_{r\ne k}Z_{r,m}\Big)\{U_{k,m}+\op(1)\},
\]
and the continuous mapping theorem, with $\Pp(\sum_{r\neq k}Z_r=0)=0$, gives ${n}^{1/2}D_n/4\dto T_K$. With equal fold sizes $\widehat\theta_+=Q_n+D_n/2$, so ${n}^{1/2}\widehat\theta_+/(2{2}^{1/2})\dto(Q^*+T_K)/{2}^{1/2}$ with $Q^*=K^{-1/2}\sum_kQ_k$ independent of all $(Z_k,U_k)$. Write $s_k=\sgn(\sum_{r\ne k}Z_r)$ and $U_k=\rho Z_k+(1-\rho^2)^{1/2}E_k$. Since the $Q_k$ are independent, symmetric about zero and independent of all the other variables, $(s_kQ_k)_k$ has, given $Z_1,\ldots,Z_K$, the distribution of $(Q_k)_k$, so
\[
\frac{Q^*+T_K}{{2}^{1/2}}\overset d=\frac1{{K}^{1/2}}\sum_ks_k\,\frac{\rho Z_k+Q_k+(1-\rho^2)^{1/2}E_k}{{2}^{1/2}}\overset d=\widetilde T_K,
\]
because $(Q_k+(1-\rho^2)^{1/2}E_k)/{2}^{1/2}$ is normal with variance $1-\rho^2/2$ and independent of $Z_1,\ldots,Z_K$. Convergence of second moments follows as in Step~5 of \S~\ref{app:counterproof}, since the estimate from each fold is exactly normal given the data of the other folds.

\textit{Variance.} Write $s_k=\sgn(S_k)$ with $S_k=\sum_{r\ne k}Z_r$, and $U_k=\rho Z_k+\sigma E_k$ with $\sigma=(1-\rho^2)^{1/2}$ and independent standard normal $E_k$. Since $s_k$ is independent of $U_k$, each $s_kU_k$ has mean zero and variance one. For $j\ne k$, terms that contain $E_j$ or $E_k$ have mean zero, so $\Cov(s_jU_j,s_kU_k)=\rho^2\E[Z_jZ_ks_js_k]$. Replace each sign by $\tanh(\cdot/\epsilon)$. The variable $Z_j$ enters $S_k$ but not $S_j$, and $Z_k$ enters $S_j$ but not $S_k$; Stein's identity, applied first in $Z_j$ and then in $Z_k$, turns $\E[Z_jZ_k\tanh(S_j/\epsilon)\tanh(S_k/\epsilon)]$ into $\epsilon^{-2}\E[\tanh'(S_j/\epsilon)\tanh'(S_k/\epsilon)]$. As $\epsilon\to0$ this tends to $4f_{jk}(0,0)$, where $f_{jk}$ is the density of $(S_j,S_k)$. The pair has variances $K-1$ and covariance $K-2$, so $4f_{jk}(0,0)=2/\{\pi((K-1)^2-(K-2)^2)^{1/2}\}=2/(\pi(2K-3)^{1/2})$. Summing over the $K(K-1)$ ordered pairs and dividing by $K$ gives $\Var\{T_K(\rho)\}=1+2\rho^2(K-1)/(\pi(2K-3)^{1/2})$.

\subsection{Modes of the limit}\label{app:shapeproof}

Table~\ref{tab:modes} locates the modes of $T_{K,d}=\rho B_{K,d}+\sigma W$ of \S~\ref{sec:directions}, which equals $T_K(\rho)$ for $d=1$. Its density is the expectation of $\sigma^{-1}\phi\{(x-\rho B_{K,d})/\sigma\}$, which we average over simulated values of $B_{K,d}$ and evaluate on a fine grid; the table reports every local maximum. For $K=2$ and $d=1$, the exact density~\eqref{eq:density2} gives the same modes, and on a grid of $\rho$ it has two modes for $\rho\ge0.75$ and one for $\rho\le0.74$, in agreement with $\rho_*=0.748$. Simulated values of $T_2(\rho)$ at three values of $\rho$ agreed with the characteristic function~\eqref{eq:cf2} and the density~\eqref{eq:density2} to within Monte Carlo error.

\begin{table}[htbp]
\centering\small
\caption{Skewness of $B_{K,d}$ and modes of the density of $T_{K,d}$ for $K$ folds and $d$ covariates}\label{tab:modes}
\begin{tabular}{rrrll}
\toprule
$d$ & $K$ & Skewness of $B_{K,d}$ & Modes, $\rho=(2/\pi)^{1/2}$ & Modes, $\rho=1/{\pi}^{1/2}$\\
\midrule
1 & 2 & 0.00 & $-0.59$, $0.59$ & $0.00$\\
1 & 3 & 0.00 & $-0.77$, $0.77$ & $0.00$\\
1 & 5 & $-0.18$ & $-0.89$, $0.82$ & $0.11$\\
1 & 10 & $-0.52$ & $-1.11$, $0.81$ & $0.33$\\
1 & 50 & $-1.59$ & $-3.75$, $0.70$ & $0.47$\\
2 & 2 & 0.00 & $0.00$ & $0.00$\\
2 & 5 & $-0.36$ & $0.47$ & $0.14$\\
2 & 10 & $-0.74$ & $0.46$ & $0.22$\\
3 & 2 & 0.00 & $0.00$ & $0.00$\\
3 & 5 & $-0.39$ & $0.32$ & $0.12$\\
3 & 10 & $-0.71$ & $0.34$ & $0.16$\\
16 & 2 & 0.00 & $0.00$ & $0.00$\\
16 & 5 & $-0.23$ & $0.11$ & $0.05$\\
16 & 10 & $-0.31$ & $0.14$ & $0.06$\\
\bottomrule
\end{tabular}
\end{table}

\subsection{Proof of Proposition~\ref{prop:dim}}\label{app:dimproof}

Throughout, $X_i\sim\N(0,I_d)$ and $\varepsilon_i\sim\N(0,1)$ are independent, $Y_i(1)=Y_i(0)=\varepsilon_i$, and each of $K$ folds of size $m$ has half of its units treated, so $\tilde Y_i=2(2A_i-1)\varepsilon_i$. Fold $k$ is split at its empirical median by the score $x\trans\widehat\beta_k$, with $\widehat\beta_k=(\sum_{i\notin I_k}X_iX_i\trans)^{-1}\sum_{i\notin I_k}X_i\tilde Y_i$.

\textit{The limit.} Put $\xi_i=(2A_i-1)\varepsilon_i$, which are independent $\N(0,1)$ variables, independent of the $X_i$ for any assignment. Let $Z_{k,m}=m^{-1/2}\sum_{i\in I_k}X_i\xi_i$ and, for a unit vector $b$ and $t\in\R$, $U_{k,m}(b,t)=m^{-1/2}\sum_{i\in I_k}\sgn(X_i\trans b-t)\xi_i$. Then ${n}^{1/2}D_n/4=K^{-1/2}\sum_kU_{k,m}(\widehat b_k,\widehat t_k)+\op(1)$, with $\widehat b_k=\widehat\beta_k/|\widehat\beta_k|$ and $\widehat t_k$ the fold median of $X\trans\widehat b_k$; the remainder accounts for the unit at the fold median, if there is one. The functions $\xi x_l$ and $\xi\,\sgn(x\trans b-t)$ form a Vapnik--Chervonenkis class with square-integrable envelope, so the fold processes converge jointly to independent Gaussian processes $(Z_k,U_k)$ with
\[
\Cov\{Z_k,U_k(b)\}=\E[X\sgn(X\trans b)]=\rho b,\qquad \Cov\{U_k(b),U_k(c)\}=\frac2\pi\arcsin(b\trans c),
\]
with $U_k(b)=U_k(b,0)$ and $\rho=(2/\pi)^{1/2}$. By the Glivenko--Cantelli theorem over half-spaces, $\sup_b|\mathrm{med}_k(X\trans b)|$ converges to zero in probability, and
\[
\E[\{\sgn(X\trans b-t)-\sgn(X\trans b)\}^2]\le4\Pp(|X\trans b|\le|t|)\to0
\]
uniformly in $b$ as $t\to0$, so stochastic equicontinuity gives $U_{k,m}(\widehat b_k,\widehat t_k)=U_{k,m}(\widehat b_k,0)+\op(1)$. Since $\E(\xi\mid X)=0$, changing the threshold does not change the mean. The empirical second-moment matrix of the training folds tends to $I_d$, so $\widehat b_k$ has the same limit as $S_{k,m}/|S_{k,m}|$, with $S_{k,m}=\sum_{r\ne k}Z_{r,m}$. The limit processes have continuous paths and $\Pp(S_k=0)=0$, so the continuous mapping theorem gives ${n}^{1/2}D_n/4\dto K^{-1/2}\sum_kU_k(b_k)$.

Decompose $U_k(b)=\rho b\trans Z_k+E_k(b)$. By the covariances above, $E_k$ is independent of $Z_k$, with $\Var E_k(b)=1-\rho^2$ for each unit $b$. The $E_k$ are independent of each other and of all $Z_r$, and each $b_k$ depends on the $Z_r$ only. Given the $Z_r$, therefore, $K^{-1/2}\sum_kE_k(b_k)\sim\N(0,1-\rho^2)$, which gives $T_{K,d}$. For the second moments, given the other folds and the covariates of fold $k$, $U_{k,m}(\widehat b_k,\widehat t_k)$ is normal with variance at most one, so its fourth moments are bounded and second moments converge.

\textit{The variance and~\eqref{eq:gamma}.} $Z_k$ is independent of $b_k$ and $b_k$ is a unit vector, so each $Z_k\trans b_k$ is exactly $\N(0,1)$. By symmetry $\E B_{K,d}=0$ and $\Var(B_{K,d})=1+(K-1)\gamma_{K,d}$, which gives the variance. For~\eqref{eq:gamma}, let $d\ge2$ and smooth the direction: $b^\epsilon(s)=s/(|s|^2+\epsilon^2)^{1/2}$, with symmetric Jacobian $J^\epsilon(s)$ satisfying $\|J^\epsilon(s)\|\le1/|s|$. Write $S_1=Z_2+R$ and $S_2=Z_1+R$ with $R=\sum_{r\ge3}Z_r$. Gaussian integration by parts in $Z_1$, on which $b^\epsilon(S_1)$ does not depend, and then in $Z_2$, on which $S_2$ does not depend, gives
\[
\E\big[(Z_1\trans b^\epsilon(S_1))(Z_2\trans b^\epsilon(S_2))\big]=\E\big[Z_2\trans J^\epsilon(S_2)b^\epsilon(S_1)\big]=\E\,\mathrm{tr}\big\{J^\epsilon(S_2)J^\epsilon(S_1)\big\}.
\]
Let $\epsilon\to0$. The left side converges by dominated convergence. On the right, $J^\epsilon(s)\to(I-bb\trans)/|s|$ for $s\ne0$, the trace is at most $d/(|S_1||S_2|)$, and this is integrable because $(S_1,S_2)$ has a bounded density on $\R^{2d}$ and $1/|s|$ is locally integrable on $\R^d$ for $d\ge2$. Finally $\mathrm{tr}\{(I-b_1b_1\trans)(I-b_2b_2\trans)\}=d-2+(b_1\trans b_2)^2$, which is positive almost surely.

\textit{Part (iii).} For $d\ge3$ the numerator of~\eqref{eq:gamma} is at most $d-1$, and by Cauchy--Schwarz $\E[1/(|S_1||S_2|)]\le\E[1/|S_1|^2]=1/\{(K-1)(d-2)\}$, since $|S_1|^2/(K-1)\sim\chi^2_d$. This is the bound. For the limit, $S_1-S_2=Z_2-Z_1$ does not grow with $K$, so $(S_1,S_2)/(K-1)^{1/2}$ converges in law to $(N,N)$ with $N\sim\N(0,I_d)$ and $b_1\trans b_2\pto1$. Thus the integrand of~\eqref{eq:gamma} times $K-1$ converges in law to $(d-1)/|N|^2$. It is uniformly integrable, since for $2(1+\delta)<d$ the same Cauchy--Schwarz step bounds its $(1+\delta)$th moment by $(d-1)^{1+\delta}\E[(\chi_d^2)^{-1-\delta}]$, free of $K$. Hence $(K-1)\gamma_{K,d}\to(d-1)\E|N|^{-2}=(d-1)/(d-2)$.

\textit{Part (ii).} With $Z_i=R_iu_i$ in polar form and $C=u_1\trans u_2$, independent of $(R_1,R_2)$, the definition gives
\begin{equation}\label{eq:two_fold_product}
B_{2,d}=\frac{Z_1\trans Z_2}{{2}^{1/2}}\Big(\frac1{R_1}+\frac1{R_2}\Big)=\frac{C(R_1+R_2)}{{2}^{1/2}},
\end{equation}
which at $d=1$ is $B_2$ of~\eqref{eq:mixture}. Write $\mu_d=\E\chi_d$, so $a_d=\mu_d^2/d$. Use $\E C^2=1/d$, $\E C^4=3/\{d(d+2)\}$, $\E R^2=d$, $\E R^3=(d+1)\mu_d$ and $\E R^4=d(d+2)$. Then $\Var(B_{2,d})=1+a_d$, so $\gamma_{2,d}=a_d$, and $\E B_{2,d}^4=3\{2d+1+2(d+1)a_d\}/(d+2)$. Subtracting $3(1+a_d)^2$ and multiplying by $\rho^4$ gives the stated $\kappa_4(T_{2,d})$. At $d=1$ it is $-(4/\pi+12/\pi^2)\rho^4$, as in~\eqref{eq:k4}.

Log-convexity of $\Gamma$ at $\frac d2,\frac{d+1}2,\frac d2+1$ gives $\mu_d^2<d$; at $\frac{d+1}2,\frac d2+1,\frac{d+3}2$ it gives $\mu_d^2\ge d^2/(d+1)$. So $d/(d+1)\le a_d<1$. For the sign, $(d+2)a_d^2+2a_d\ge d(d^2+4d+2)/(d+1)^2>d-1$, because $d^3+4d^2+2d>(d-1)(d+1)^2$.

\textit{Part (i).} By~\eqref{eq:gamma} for $d\ge2$ and Theorem~\ref{thm:counter} for $d=1$, $\sigma^2:=\Var(T_{K,d})>1$. Suppose that $T_{K,d}\sim\N(0,\sigma^2)$. By Cauchy--Schwarz, $|B_{K,d}|\le K^{-1/2}\sum_k|Z_k|\le|Z|$, where $Z\in\R^{Kd}$ stacks the $Z_k$. Since $|Z|$ is a 1-Lipschitz function of a standard Gaussian vector and $\E|Z|\le{Kd}^{1/2}$, Gaussian concentration gives $\E e^{s|Z|}\le\exp(s{Kd}^{1/2}+s^2/2)$ for $s\ge0$. Hence, for every $t$,
\[
e^{t^2\sigma^2/2}=\E e^{tT_{K,d}}=e^{t^2(1-\rho^2)/2}\,\E e^{t\rho B_{K,d}}\le\exp\big(|t\rho|{Kd}^{1/2}+t^2/2\big),
\]
which fails for large $|t|$. A rescaled normal variable is normal, so no rescaling helps.

\textit{Part (iv).} Let $A_{jr}=Z_j\trans Z_r/{d}^{1/2}$ for $j<r$. It is a normalized sum over coordinates of independent and identically distributed vectors $(Z_{jl}Z_{rl})_{j<r}$ with mean zero and identity covariance, so the $A_{jr}$ converge jointly to independent $\N(0,1)$. Also $|S_k|^2/d\pto K-1$. Therefore $Z_k\trans b_k=(K-1)^{-1/2}\sum_{r\ne k}A_{kr}+\op(1)$, and
\[
B_{K,d}=\frac{2}{(K(K-1))^{1/2}}\sum_{j<r}A_{jr}+\op(1)\dto\N(0,2).
\]
Each $Z_k\trans b_k$ is exactly standard normal, so $B_{K,d}$ has bounded fourth moments, $\Var(B_{K,d})\to2$, and $(K-1)\gamma_{K,d}\to1$.

\textit{Numerical values.} Tables~\ref{tab:dimtwo} and~\ref{tab:dimbound} give numerical values of the quantities in the proposition.

\begin{table}[htbp]
\centering\small
\caption{Variance, excess kurtosis and rejection rates (\protect\%) of nominal 5\protect\% tests for the two-fold limit with $d$ covariates and $\rho^2=2/\pi$}\label{tab:dimtwo}
\begin{tabular}{rccccc}
\toprule
$d$ & $a_d$ & $\Var(T)$ & excess kurtosis & normal & normal scaled by exact SD\\
\midrule
1 & 0.637 & 1.405 & $-0.511$ & 8.8 & 3.75\\
2 & 0.785 & 1.500 & $-0.410$ & 10.5 & 4.10\\
3 & 0.849 & 1.540 & $-0.338$ & 11.3 & 4.33\\
4 & 0.884 & 1.562 & $-0.286$ & 11.7 & 4.48\\
8 & 0.940 & 1.598 & $-0.176$ & 12.3 & 4.74\\
16 & 0.969 & 1.617 & $-0.099$ & 12.4 & 4.88\\
64 & 0.992 & 1.632 & $-0.027$ & 12.5 & 4.98\\
$\infty$ & 1 & 1.637 & 0 & 12.6 & 5\\
\bottomrule
\end{tabular}
\end{table}

\begin{table}[htbp]
\centering\small
\caption{Values of $(K-1)\gamma_{K,d}$, the excess of the variance of $B_{K,d}$ over one, for $K$ folds and $d$ covariates}\label{tab:dimbound}
\begin{tabular}{rcccccc}
\toprule
$d$ & $K=2$ & $5$ & $10$ & $50$ & $200$ & $K\to\infty$\\
\midrule
1 & 0.637 & 0.962 & 1.390 & 3.167 & 6.36 & $\infty$\\
2 & 0.788 & 1.052 & 1.324 & 2.032 & 2.694 & \\
3 & 0.850 & 1.062 & 1.247 & 1.592 & 1.769 & 2\\
4 & 0.884 & 1.060 & 1.190 & 1.382 & 1.458 & 1.5\\
8 & 0.940 & 1.037 & 1.094 & 1.151 & 1.163 & 1.167\\
16 & 0.969 & 1.021 & 1.045 & 1.066 & 1.070 & 1.071\\
$\infty$ & 1 & 1 & 1 & 1 & 1 & 1\\
\bottomrule
\end{tabular}
\end{table}

Simulation agrees with the formulae; for instance, $\Var(B_{5,4})$ is $2.058$ from~\eqref{eq:gamma} and $2.060$ by simulation. For $K\ge3$ the limit is also skewed to the left, with skewness $-0.38$ for $B_{5,4}$ and $-0.89$ for $B_{20,4}$; the skewness vanishes as $d$ grows.

\subsection{Published conditions}\label{app:published}

This subsection sets out the conditions of Theorem~6 of \citet{imai2025} and gives the lemma used in \S~\ref{sec:published}. Table~\ref{tab:assumptions} lists each condition as stated in that paper and the corresponding fact in the model of \S~\ref{sec:univariate}. Their estimator of the effect in group $k$ of fold $\ell$, with $K$ groups, $m_1$ treated and $m_0$ control units in the fold and group indicator $\hat f_k^\ell$, is $K\{m_1^{-1}\sum_iY_iT_i\hat f_k^\ell(X_i)-m_0^{-1}\sum_iY_i(1-T_i)\hat f_k^\ell(X_i)\}$; with two groups and $p=m_1/m$ this is $\hat\theta_{k,\pm}=2P_k(\hat w_{k,\pm}\tilde Y)$ of \S~\ref{app:notation}. Their target is the average of~\eqref{eq:target} over training samples, and their theorem standardizes by the exact variance of the estimator.

\begin{table}[htbp]
\centering\small
\caption{Conditions of Theorem~6 of \citet{imai2025} and the corresponding facts in the model of \S~\ref{sec:univariate}}\label{tab:assumptions}
\begin{tabular}{p{4.3cm}p{9.2cm}}
\toprule
Condition & Fact in the model\\
\midrule
Assumption 1, no interference & Each outcome depends only on the unit's own treatment\\
Assumption 2, random sampling & Potential outcomes and covariates are independent and identically distributed\\
Assumption 3, complete randomization & Half of the units are treated by complete randomization\\
Folds & Their \S~3.1 takes folds of equal size with the same numbers of treated and control units; here each arm is split at random into $K$ folds\\
Bijective score & Assumed there without loss of generality; the learned slope is non-zero with probability one, so the score $\hat\beta x$ is one-to-one\\
Assumption 4 & Not a condition of Theorem~6, which replaces it by Assumption~6, its version for cross-fitting\\
Assumption 5, moments & Positive variance and finite third moment of each potential outcome are required; both are standard normal\\
Assumption 6, continuity under cross-fitting & The effect as a function of the score quantile is identically zero for every training sample, hence continuous, of bounded variation and with bounded maximum\\
Assumption 7, stability & The mean effect in every score interval is zero for every training sample, so the norm in~\eqref{eq:ILcondition} is identically zero\\
Number of folds & Fixed, as required\\
Target and standardization & The effect in the upper half averaged over training samples, which is zero, and the exact variance of its estimator\\
\bottomrule
\end{tabular}
\end{table}

\begin{lemma}[A sufficient condition for replacing a learned ranking by a fixed one]\label{lem:weight_l2}
Under $H_\tau$, let $\nu=\tilde Y-\tau$ with $\E \nu^2<\infty$. Let the training sample $\calT$ be independent of the evaluation fold, let $g_{\calT}$ be a bounded weight determined by $\calT$, which represents the learned ranking, and let $g^0$ be a fixed bounded weight. If, for an independent new unit,
\begin{equation}\label{eq:weight_l2}
 V_{\calT}:=\E\big[(g_{\calT}(X)-g^0(X))^2\nu^2\mid\calT\big]\pto0,
\end{equation}
then ${m}^{1/2}P_k[(g_{\calT}-g^0)\nu]=\op(1)$. The conclusion holds under independent Bernoulli assignment and when the treated proportion of each fold is fixed, the expectation then being taken over the mixture of the two arms.
\end{lemma}

\begin{proof}
Write $h_{\calT}=(g_{\calT}-g^0)\nu$. Under Bernoulli assignment, $H_\tau$ gives $\E(h_{\calT}\mid\calT)=0$, and independence of training and evaluation gives $\Var({m}^{1/2}P_kh_{\calT}\mid\calT)=V_{\calT}$. With fixed arm sizes, let $p_1=p$ and $p_0=1-p$; the weighted mean over the two arms is still zero, and the conditional variance is at most $\sum_ap_a\E(h_{\calT}^2\mid A=a,\calT)=V_{\calT}$. For any $\epsilon,\eta>0$, the conditional Chebyshev inequality gives $\Pp\{|{m}^{1/2}P_kh_{\calT}|>\epsilon\}\le\Pp(V_{\calT}>\eta)+\eta/\epsilon^2$. Letting $m\to\infty$ and then $\eta\downarrow0$ proves the claim.
\end{proof}

In the model of \S~\ref{sec:univariate}, take $g^0(X)=\sgn(X)$. Then $g_{\calT}(X)=\sgn(\widehat\beta_{\calT})\sgn(X)$ and $\nu=2\xi$ with $\xi=(2A-1)\varepsilon$, so
\begin{equation}\label{eq:weight_l2_counter}
 V_{\calT}=4\{\sgn(\widehat\beta_{\calT})-1\}^2=16\,\ind\{\widehat\beta_{\calT}<0\},
\end{equation}
which equals 16 with probability $1/2$, while the left-hand side of~\eqref{eq:ILcondition} is identically zero. With $g^0=-\sgn$ the same holds on the complementary event, and any other fixed bounded weight differs from $\sgn(X)$ and from $-\sgn(X)$ on sets of positive probability, so that $V_{\calT}$ is bounded away from zero on both events. The proof of Theorem~6 in their Supplementary Material~S9 writes each contribution of the learned ranking as that of a fixed ranking plus a remainder whose root mean square is stated to be $o(m^{-1/2})$, citing Assumption~7; in this model the mean square of that remainder is of the order of $V_{\calT}$, which does not vanish for any fixed ranking. Condition~\eqref{eq:weight_l2} is thus sufficient for replacing the learned ranking by a fixed one, but it is not necessary for normality: by Proposition~\ref{prop:structure} the limit is normal whenever $\Sigma_{ZU}=0$, whether or not the ranking converges.

The assumption behind normal intervals for the cross-validated area under the receiver operating characteristic curve has the same form. For a label $Y\in\{0,1\}$ independent of $X$, $p=\Pp(Y=1)$ and a score $s$ with continuous distribution function $F_s$, the influence curve of \citet[Theorem~4.1]{ledell2015} equals $\{Y/p-(1-Y)/(1-p)\}\{F_s(s(X))-1/2\}$. Their convergence assumption is therefore~\eqref{eq:weight_l2} with $\nu=Y/p-(1-Y)/(1-p)$ and the rank weight $g_{\calT}=F_{\hat s}(\hat s(X))-1/2$ of the learned score $\hat s$.

\section{Local and strong signals}\label{sec:signallimits}

This section proves Theorem~\ref{thm:signals} under the design~\eqref{eq:signal_dgp}, with the notation~\eqref{eq:signal_t}--\eqref{eq:Rdelta}.

\begin{figure}
\figuresize{.85}
\figurebox{20pc}{25pc}{}[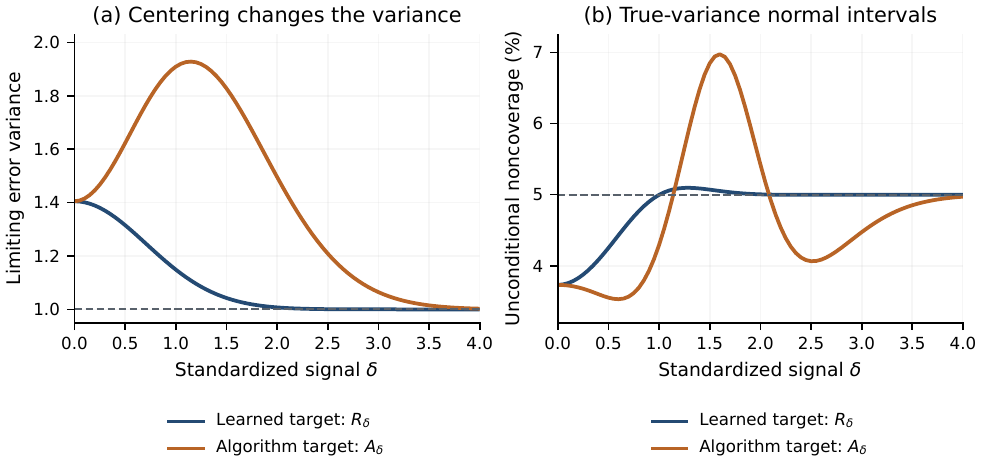]
\caption{The two-fold limit under local alternatives depends on the centring. (a) Variance of the limit. (b) Probability that a nominal 95\protect\% normal interval with the exact variance misses its target; the dashed line marks 5\protect\%.}
\label{fig:signal}
\end{figure}

The curves in Fig.~\ref{fig:signal} are computed by numerical integration over $(Z_1,Z_2)$, given which $R_\delta$ is normal with mean $\rho(s_2^\delta Z_1+s_1^\delta Z_2)/{2}^{1/2}$ and variance $1-\rho^2$. At $\delta=1$ the limiting variance is $1.15$ when the estimate is centred at the group difference $\Delta^{\rm learn}_m$ of the learned rankings and $1.91$ when it is centred at its average $\Delta^{\rm alg}_m$ over training samples; at $\delta=1.5$, normal intervals with these variances miss their targets $5.1\%$ and $6.9\%$ of the time. In simulations of the full procedure with $n=4000$ and $\delta=1$, the normal interval for $\Delta^{\rm alg}_m$ missed it $4.3\%$ of the time, and the interval with the critical value of the limit $A_\delta$ missed it $5.1\%$ of the time.

\textit{The probability of a wrong sign and the effect of the empirical median.}
Given the covariates of a fold, $\sum X_i\xi_i/(\sum X_i^2)^{1/2}$ is standard normal with a conditional distribution that does not depend on $X$, so it is independent of $\sum X_i^2\sim\chi_m^2$. Hence
\[
 \widehat\beta_k
 =\theta_m+2\frac{\sum X_i\xi_i}{\sum X_i^2},
 \qquad
 \frac{{m}^{1/2}}{2}(\widehat\beta_k-\theta_m)\sim t_m.
\]
This gives~\eqref{eq:signal_t} and, for $\theta_m\neq0$, the exact probability $q_m=F_{t_m}(-|\delta_m|)$ that the slope learned on a fold has the wrong sign. In the population, $X$ has mean $c$ in the upper half of the ranking when the sign is correct and $-c$ when it is wrong, which gives~\eqref{eq:signal_targets}.

Let $\widehat q_k$ be the average of the two middle order statistics of fold $k$ and $r_k$ the empirical median-split weight. The weights change for $C_k=|N_k^+-m/2|=\Op({m}^{1/2})$ units, all with $|X_i|\le|\widehat q_k|$. The median of a normal sample satisfies $\widehat q_k=\Op(m^{-1/2})$. Therefore
\begin{equation}\label{eq:median_signal}
 \left|P_k(r_kX)-P_k|X|\right|
 \le\frac{2C_k|\widehat q_k|}{m}=\Op(m^{-1}).
\end{equation}
The noise part follows~\eqref{eq:medianbound}: ${m}^{1/2}P_k[(r_k-\sgn(X))\xi]=\op(1)$, with the same $L^2$ bound. For bounded $\theta_m$,
\begin{equation}\label{eq:signal_rank_expansion}
 P_k(r_k\tilde Y)
 =\theta_m P_k|X|+2P_k\{\sgn(X)\xi\}+\op(m^{-1/2}).
\end{equation}
The empirical median has a negligible effect on the signal term because $\tau_m(x)=\theta_mx$ vanishes at the population median $x=0$, not because the conditional average effect is constant.

\textit{Proof of Theorem~\ref{thm:signals}(i).}
Let $Z_{k,m}=m^{-1/2}\sum_{i\in I_k} X_i\xi_i$ and $U_{k,m}=m^{-1/2}\sum_{i\in I_k}\sgn(X_i)\xi_i$ be the moments of the noise, which have the joint normal limit obtained in the proof of Theorem~\ref{thm:counter}. The training coefficients satisfy
\[
 \frac{{m}^{1/2}}{2}\widehat\beta_k
 =\delta_m+\frac{Z_{k,m}}{P_kX^2}
 =\delta_m+Z_{k,m}+\op(1).
\]
Since $\Pp(Z_k+\delta=0)=0$, the signs of the learned slopes converge jointly in distribution to $s_k^\delta$. By~\eqref{eq:signal_rank_expansion} and $P_k|X|\pto c$,
\[
 \frac{{m}^{1/2}}{2}P_k(r_k\tilde Y)
 =U_{k,m}+\delta_m P_k|X|+\op(1)
 =U_{k,m}+\delta c+\op(1).
\]
Substituting into $D_n=\sum_ks_{3-k}P_k(r_k\tilde Y)$ gives the uncentred limit
\[
 \frac{{n}^{1/2}D_n}{4}\dto \frac{s_2^\delta(U_1+\delta c)+s_1^\delta(U_2+\delta c)}{{2}^{1/2}}.
\]
Moreover,
\[
 \frac{{n}^{1/2}\Delta^{\mathrm{learn}}_m}{4}
 =\frac{\delta_mc}{{2}^{1/2}}(s_1+s_2),\qquad
 \frac{{n}^{1/2}\Delta^{\mathrm{alg}}_m}{4}
 ={2}^{1/2}\,\delta_mc\,\bar s_m\longrightarrow{2}^{1/2}\,\delta c\,\bar s,
\]
where $\bar s_m\to\bar s$ follows from the convergence of the $t_m$ distribution and $\delta_m\to\delta$. Subtracting gives the convergence in Theorem~\ref{thm:signals}(i).

For second moments, ${m}^{1/2}\theta_m$ is bounded and $|P_k(r_kX)|\le P_k|X|$ has uniformly bounded moments of every fixed order. Given the covariates of fold $k$, ${m}^{1/2}P_k(r_k\xi)$ is exactly standard normal, and multiplication by the other fold's sign leaves its absolute moments unchanged. The triangle inequality bounds the fourth moments of the standardized estimator uniformly, and $n^{1/2}\Delta^{\mathrm{learn}}_m$ and $n^{1/2}\Delta^{\mathrm{alg}}_m$ are bounded, so the squares of the centred and scaled estimates are uniformly integrable. Using $\E s_k^\delta=\bar s$, $\E U_k=0$ and $\E(s_k^\delta U_k)=2\rho\phi(\delta)=\gamma$,
\[
 \E R_\delta=0,\quad \Var(R_\delta)=1+\gamma^2,\quad
 \Var\{(s_1^\delta+s_2^\delta)/{2}^{1/2}\}=1-\bar s^2,
\]
and $\Cov\{R_\delta,(s_1^\delta+s_2^\delta)/{2}^{1/2}\}=\bar s\gamma$, which gives~\eqref{eq:varlocala}.

To see that the limits are not normal, take a Gaussian pair $(Z,U)$ with standard marginals and correlation $\rho$. Completing the square in the bivariate normal density gives, for $a\in\{-1,1\}$,
\begin{equation}\label{eq:tilt}
 \E\{e^{vU}\ind(\sgn(Z+\delta)=a)\}
 =e^{v^2/2}\Phi(a\delta+a\rho v).
\end{equation}
Expanding over the four sign combinations, with $v=\rho t/{2}^{1/2}$,
\begin{equation}\label{eq:mgfr}
 e^{-t^2/2}\E e^{tR_\delta}
 =\Phi(\delta+v)^2+\Phi(-\delta+v)^2+2\Phi(\delta-v)\Phi(-\delta-v)
 =1+\{\Phi(\delta+v)-\Phi(\delta-v)\}^2 .
\end{equation}
For $\rho\neq0$ and finite $\delta$, the variance of $R_\delta$ is $1+4\rho^2\phi(\delta)^2>1$, while the ratio in~\eqref{eq:mgfr} lies in $(1,2]$ for all $t\neq0$; for a centred normal variable with this variance the same ratio is unbounded as $|t|\to\infty$. So $R_\delta$ is not normal. The same expansion, keeping the term in $s_1^\delta+s_2^\delta$, gives, with $\kappa=\delta c$,
\[
 \E e^{tA_\delta}
 =e^{t^2/2-{2}^{1/2}\kappa\bar st}\big\{
 e^{{2}^{1/2}\kappa t}\Phi(\delta+v)^2
 +e^{-{2}^{1/2}\kappa t}\Phi(-\delta+v)^2
 +2\Phi(\delta-v)\Phi(-\delta-v)\big\}.
\]
In the model of \S~\ref{sec:signal}, $\rho=c>0$ and $\Var(A_\delta)>1$. For $\delta>0$, as $t\to+\infty$, $\log\E e^{tA_\delta}=t^2/2+{2}^{1/2}\kappa(1-\bar s)t+o(1)$, whose quadratic coefficient $1/2$ cannot match a centred normal law with variance larger than one. At $\delta=0$, $A_0=R_0$, and the map $(\delta,Z_k,U_k)\mapsto(-\delta,-Z_k,-U_k)$ leaves the law of $A_\delta$ unchanged, which covers $\delta<0$. Since second moments converge and the limiting variances are positive, standardization by the exact variance does not change this conclusion.

\textit{Proof of Theorem~\ref{thm:signals}(ii).}
Let $s_m^*=\sgn(\theta_m)$, and let $D_n^*$ be the group difference computed from the same data with both learned signs replaced by $s_m^*$. Both learned signs equal $s_m^*$ with probability $(1-q_m)^2\to1$, and on that event $D_n=D_n^*$ and $\Delta^{\mathrm{learn}}_m=2|\theta_m|c$. Moreover,
\[
 {n}^{1/2}\,|\Delta^{\mathrm{alg}}_m-2|\theta_m|c|
 =4{n}^{1/2}\,|\theta_m|c q_m\longrightarrow0,
\]
because for $m>2$, $q_m\le \E(t_m^2)/\delta_m^2=m/\{(m-2)\delta_m^2\}$, so $|\delta_m|q_m\to0$. Both $\Delta^{\mathrm{learn}}_m$ and $\Delta^{\mathrm{alg}}_m$ therefore differ from $2|\theta_m|c$ by $\op(n^{-1/2})$. By~\eqref{eq:signal_rank_expansion},
\begin{equation}\label{eq:ifd}
 {n}^{1/2}(D_n^*-2|\theta_m|c)
 =\frac1{{n}^{1/2}}\sum_{i=1}^n
 \left\{2|\theta_m|(|X_i|-c)
       +4s_m^*\sgn(X_i)\xi_i\right\}+\op(1).
\end{equation}
Since $\Var(|X|)=1-c^2$ and the cross term linear in $\xi$ has mean zero, the influence function has variance $16+4\theta_m^2(1-c^2)$. Because $\theta_m$ is bounded, the summands have uniformly bounded fourth moments, so the Lindeberg condition holds and Theorem~\ref{thm:signals}(ii) follows. The argument remains valid when $\theta_*=0$ and the sign of $\theta_m$ varies with $m$.

\textit{Proof of Theorem~\ref{thm:signals}(iii).}
Conditionally on the assignment, the data are equivalent to independent pairs with $X\sim\N(0,1)$ and $\tilde Y\mid X\sim\N(2\delta X/{m}^{1/2},4)$, and the distribution of the assignment does not depend on $\delta$. The Hellinger affinity of two normal densities with common variance is the exponential of minus the squared mean difference divided by eight times the variance. With $d=\delta_1-\delta_0$, integrating over one unit gives
\[
 \E_X\exp\{-d^2X^2/(8m)\}
 =\{1+d^2/(4m)\}^{-1/2},
\]
and the product over $n=2m$ independent units gives~\eqref{eq:affinity}. For densities $p,q$, the Cauchy--Schwarz inequality gives
\[
 d_{\mathrm{TV}}(P,Q)
 =\frac12\int|{p}^{1/2}-{q}^{1/2}|({p}^{1/2}+{q}^{1/2})
 \le(1-\mathcal A(P,Q)^2)^{1/2},
\]
so the limit superior of the total variation distance between the distributions of the data at $\delta_0$ and $\delta_1$ is strictly below one. An estimator consistent at both $\delta_0$ and $\delta_1$ would yield a test whose two error probabilities both tend to zero, which forces the total variation distance to tend to one, a contradiction.

\section{Joint calibration}\label{app:calibration}

Algorithm~\ref{alg:calibration} states the procedure of \S~\ref{sec:calibration}.

\begin{algorithm}[htbp]
\caption{Joint calibration of a cross-fitted test}\label{alg:calibration}
\KwIn{folds $I_1,\ldots,I_K$; training sets $F(k)$; features $h$; candidate rankings $g_1,\ldots,g_J$; selection function $\omega$; number of draws $N$.}
Compute $Z_{k,n}$ and $U_{k,n}$ $(k=1,\ldots,K)$, with weights centred at empirical medians\;
Estimate $\Sigma$ by the sample covariance of the unit-level vectors $V_i$, with $\tau$ replaced by $Q_n$; when the arm sizes of the folds are fixed, centre within arms and weight the arms by their shares (\S~\ref{sec:design})\;
Compute $T_n$ from \eqref{eq:Tn}\;
\For{$b=1,\ldots,N$}{
Draw independent vectors $(Z_k^{*\T},U_k^{*\T})^\T\sim N(0,\hat\Sigma)$ $(k=1,\ldots,K)$\;
Compute $T_b^*$ from \eqref{eq:Tn}, with $(Z_k^*,U_k^*)$ in place of $(Z_{k,n},U_{k,n})$ and $\hat\Sigma_{UU}$ unchanged\;
}
\KwOut{symmetric $p$-value $\hat p=\{1+\#(b:|T_b^*|\geq|T_n|)\}/(N+1)$; equal-tailed $p$-value $\hat p_{\rm eq}=\min\{1,2\min(\hat p_-,\hat p_+)\}$, where $\hat p_-=\{1+\#(b:T_b^*\leq T_n)\}/(N+1)$ and $\hat p_+=\{1+\#(b:T_b^*\geq T_n)\}/(N+1)$.}
\end{algorithm}

\subsection{Empirical quantiles and ties}\label{sec:quantile}

For each fixed score $s_j$, assume its population median $q_j$ satisfies one of two conditions:
(i) the score distribution has no atom at $q_j$, and for every $\delta>0$, $\Pp(s_j\le q_j-\delta)<1/2<\Pp(s_j\le q_j+\delta)$; or
(ii) $q_j$ lies strictly inside an atom, $\Pp(s_j>q_j)<1/2<\Pp(s_j\ge q_j)$.
In case (i) let $w_j(x)=\ind\{s_j(x)>q_j\}$. In case (ii) let
\begin{equation}\label{eq:tie}
 w_j(x)=\ind\{s_j(x)>q_j\}+\lambda_j\ind\{s_j(x)=q_j\},\qquad \lambda_j=\frac{1/2-\Pp(s_j>q_j)}{\Pp(s_j=q_j)}\in(0,1).
\end{equation}
Let $g_j=2w_j-1$. The sample versions replace probabilities by fold frequencies and the median by the empirical median, so that $P_k\widehat g_{k,j}=0$ exactly.

\begin{lemma}[First-order expansion of the sample moments]\label{lem:quantile}
Under $H_\tau$,~\eqref{eq:moments_general} and the quantile conditions above, the moments~\eqref{eq:ZUhat} satisfy~\eqref{eq:expansions}. The sample covariance of the unit-level vectors is consistent for $\Sigma$ whenever $\widehat\tau\pto\tau$. A median on the boundary of an atom satisfies neither condition and is not covered by the lemma.
\end{lemma}

The idea of the proof is that, because $\E(\nu\mid X)=0$, replacing the population median by the empirical median does not change the mean of an evaluation moment, and the second moment of the contributions of the units whose weights change tends to zero. Only consistency of the empirical median is needed, not a rate of convergence.

\textit{Proof of Lemma~\ref{lem:quantile}.}\label{app:primitive} We work under independent Bernoulli assignment. Write $\mu_h=E h(X)$, $\overline h_k=P_kh$ and $\overline\nu_k=P_k\nu$. Sample centring gives the exact decomposition
\begin{equation}\label{eq:trainingexpand}
 {m}^{1/2}P_k[(h-\overline h_k)\tilde Y]
 ={m}^{1/2}P_k[(h-\mu_h)\nu]
 -{m}^{1/2}(\overline h_k-\mu_h)\overline\nu_k.
\end{equation}
Finite second moments give $\overline h_k-\mu_h=\Op(m^{-1/2})$ and $\overline\nu_k=\Op(m^{-1/2})$, so the last term is $\op(1)$. If the features are centred at their mean over all the folds in $F(k)$ instead of fold by fold, the result changes by finitely many products of errors of sample means, which are $\op(1)$.

For the evaluation moments, exact group centring gives
\[
 {m}^{1/2}P_k(\widehat g_{k,j}\tilde Y)
 ={m}^{1/2}P_k(\widehat g_{k,j}\nu),
\]
and it remains to show
\begin{equation}\label{eq:evalremainder}
 {m}^{1/2}P_k[(\widehat g_{k,j}-g_j)\nu]=\op(1).
\end{equation}
Under condition (i), the empirical median satisfies $\widehat q_{k,j}\pto q_j$. Fix $\delta>0$. On the event $|\widehat q_{k,j}-q_j|\le\delta$, weights can differ only for units with $|s_j(X)-q_j|\le\delta$, and every difference is bounded by 2 in absolute value. Given all covariates of the fold, the $\nu_i$ are independent with mean zero. With $\sigma_\nu^2(X)=\E(\nu^2\mid X)$, the conditional variance is at most
\[
 4P_k\big[\ind\{|s_j(X)-q_j|\le\delta\}\sigma_\nu^2(X)\big].
\]
Letting $m\to\infty$, the law of large numbers sends the right side to its population expectation; letting $\delta\downarrow0$, the absence of an atom and $\E\nu^2<\infty$ send that expectation to zero. The conditional Chebyshev inequality gives~\eqref{eq:evalremainder}.

Under condition (ii), the law of large numbers for the frequencies of $\{s_j>q_j\}$ and $\{s_j\ge q_j\}$ gives $\Pp(\widehat q_{k,j}=q_j)\to1$ and $\widehat\lambda_{k,j}\pto\lambda_j$. On that event the remainder is
\[
 2(\widehat\lambda_{k,j}-\lambda_j)
 {m}^{1/2}P_k[\ind\{s_j(X)=q_j\}\nu]=\op(1),
\]
since the last standardized sum is $\Op(1)$. Finite $J$ and $K$ give all expansions simultaneously.

For the covariance, let $\widehat V_i$ be the vector~\eqref{eq:V} of unit $i$ with $\tau$, $\mu_h$ and $g_j$ replaced by $\widehat\tau$, $\overline h_k$ and $\widehat g_{k,j}$. Under condition (i), the law of large numbers on the neighbourhood $|s_j(X)-q_j|\le\delta$, with envelope $\nu^2$, and under condition (ii), consistency of $\widehat\lambda_{k,j}$, give
\[
 P_k\|\widehat V-V\|^2=\op(1);
\]
the terms that arise from replacing $\tau$ and $\mu_h$ are controlled by $\widehat\tau\pto\tau$, $\overline h_k\pto\mu_h$ and~\eqref{eq:moments_general}. By the Cauchy--Schwarz inequality, the sample second-moment matrices of $\widehat V$ and $V$ then differ by $\op(1)$, and the law of large numbers for $P_k(VV^\T)$, with $P_kV\pto0$, completes the proof.

\subsection{Proofs of Theorems~\ref{thm:general} and~\ref{thm:calibration}}\label{app:calibproof}

The folds are disjoint sets of independent rows, so the raw sums of different folds are independent. The expansion~\eqref{eq:expansions}, the multivariate central limit theorem and this independence give
\[
 ((Z_{1,n},U_{1,n}),\ldots,(Z_{K,n},U_{K,n}))
 \dto((Z_1,U_1),\ldots,(Z_K,U_K)),
\]
where the limits are the independent $N(0,\Sigma)$ vectors of Theorem~\ref{thm:general}. Consider the map
\[
 \mathcal F((z_1,u_1),\ldots,(z_K,u_K);\Sigma)
 =\frac{\sum_{k=1}^K\omega_k\trans u_k}{\big(\sum_{k=1}^K\omega_k\trans\Sigma_{UU}\omega_k\big)^{1/2}},\qquad \omega_k=\omega\left(\sum_{r\in F(k)}z_r\right).
\]
The sets $F(k)$ are fixed. The map is continuous in $\Sigma$, and it can be discontinuous in its other arguments only where the selection function is discontinuous at one of the $K$ sums of training moments. By assumption the limiting vector lies at such a point with probability zero, and the denominator is positive almost surely. Since $\widehat\Sigma_n\pto\Sigma$, the continuous mapping theorem gives the limit~\eqref{eq:TSigma}. Only almost-sure continuity is used, so the selection function may be discontinuous on any set to which the sums of training moments assign probability zero. When folds $j$ and $k$ train each other, $j\in F(k)$ and $k\in F(j)$, the selection $\omega(\sum_{r\in F(k)}Z_r)$ depends on $Z_j$, which is correlated with $U_j$; this dependence is kept by the map and is the source of the non-normal limit.

For the reference distribution with an estimated covariance, we use the continuity of the symmetric square root of a positive semidefinite matrix. With independent standard normal vectors $\zeta_1,\ldots,\zeta_K$, write the reference vectors as $\widehat\Sigma_n^{1/2}\zeta_k$. Every subsequence has a further subsequence along which $\widehat\Sigma_n\to\Sigma$ almost surely, and along it the reference vectors converge almost surely to $\Sigma^{1/2}\zeta_k$. The map is almost surely continuous at the limit, so for every bounded continuous test function $b$, dominated convergence gives
\[
 \E\{b(T_n^*)\mid\text{data}\}\longrightarrow\E b(T_\Sigma)
\]
almost surely. Since this holds along a further subsequence of every subsequence, the conditional distribution of $T_n^*$ converges weakly to the distribution of $T_\Sigma$, in probability. The same proof applies when a separate consistent covariance estimator is used for each fold.

For the reference distribution of the absolute value, weak convergence and the stated quantile condition give consistency of the $(1-\alpha)$ quantile. Conditionally on the data, the $\Nref$ draws are independent, so as $\Nref\to\infty$ their empirical distribution function converges uniformly, in probability, to the conditional distribution function of $|T^*|$. The empirical critical value therefore satisfies $\widehat c_{\alpha,n}\pto c_\alpha$. Finally, $|T_n|\dto|T_\Sigma|$, the absence of an atom at the critical value and Slutsky's theorem give the first statement of Theorem~\ref{thm:calibration}. For the equal-tailed region, the same argument applied to the distribution of $T_\Sigma$ itself gives consistency of the empirical $\alpha/2$ and $1-\alpha/2$ quantiles, and $T_n\dto T_\Sigma$ gives the level. If each fold is standardized separately instead of through the common denominator, only the map $\mathcal F$ changes and the argument applies as it stands.

\subsection{Fixed arm sizes}\label{sec:design}

Suppose that the units are sampled independently from a population and that treatment is assigned by complete randomization with a fixed number of treated units in each fold; take $p$ in~\eqref{eq:tildeY} to be the treated proportion. Let $V_a$ be the vector~\eqref{eq:V} for a unit in arm $a$. The covariance in the joint central limit theorem is
\begin{equation}\label{eq:designcov}
 \Sigma_{\mathrm{des}}=p\Cov(V_1)+(1-p)\Cov(V_0),
\end{equation}
estimated by centring within each arm and weighting the arms by their proportions. It is related to the covariance $\Sigma_{\mathrm{iid}}$ under independent Bernoulli assignment by
\begin{equation}\label{eq:design_difference}
 \Sigma_{\mathrm{iid}}=\Sigma_{\mathrm{des}}+p(1-p)(\E V_1-\E V_0)(\E V_1-\E V_0)\trans.
\end{equation}
The expansion~\eqref{eq:expansions} still holds under $H_\tau$, because the means of the two arms, weighted by their proportions, cancel, as shown below. Theorem~\ref{thm:calibration} therefore holds with $\Sigma_{\mathrm{des}}$ in place of $\Sigma$. For cluster-randomized or stratified experiments, $\Sigma$ must be replaced by the limiting covariance under the design actually used.

\textit{Proof.}\label{app:design} Let the treated share of every fold equal a fixed $p$, with $m_a$ units in arm $a$ of each fold and $m_1/m=p$. Given the assignment and the split into folds, and after relabelling, the two arms are independent samples from their respective populations. Write $P_{k,a}$ for the mean within arm $a$ of fold $k$, $P_a$ for the population distribution of that arm, $\tilde Y_a$ and $\nu_a=\tilde Y_a-\tau$ for the transformed outcome and its centred version in arm $a$, and $p_1=p$, $p_0=1-p$. Under $H_\tau$,
\[
 \sum_{a=0}^1p_aP_aV_a=0,
\]
so
\begin{equation}\label{eq:armclt}
 \frac1{{m}^{1/2}}\sum_{a=0}^1\sum_{i\in I_k:A_i=a}V_{a,i}
 =\sum_{a=0}^1(p_a)^{1/2}\,
 \frac1{(m_a)^{1/2}}\sum_{i\in I_k:A_i=a}(V_{a,i}-P_aV_a).
\end{equation}
Independent central limit theorems within the arms give~\eqref{eq:designcov}, and the population decomposition of the covariance gives~\eqref{eq:design_difference}. Consistency of the within-arm sample covariance follows from the law of large numbers for second moments in each arm and the plug-in control in the proof of Lemma~\ref{lem:quantile}.

Within an arm, $\E(\nu_a\mid X)$ is not zero, so the argument of Lemma~\ref{lem:quantile} cannot be applied arm by arm. The means cancel only when the two arms are combined: for any bounded weight $v$ that depends only on $X$, randomization and $H_\tau$ give
\begin{equation}\label{eq:driftcancel}
 \sum_{a=0}^1p_aP_a[v(X)\nu_a]
 =\E[v(X)\{\tau(X)-\tau\}]=0.
\end{equation}
Identity~\eqref{eq:driftcancel} holds for every such weight, and hence for the data-dependent weight $v=\widehat g-g$, so replacing the population median by the empirical one does not change the mean. What remains is a sum of empirical processes centred within each arm:
\[
 {m}^{1/2}P_k[(\widehat g-g)\nu]
 ={m}^{1/2}\sum_a p_a(P_{k,a}-P_a)[(\widehat g-g)\nu_a].
\]
Under condition (i) of \S~\ref{sec:quantile}, $\{\nu_a\ind(s_j\le t):t\in\R\}$ is a Vapnik--Chervonenkis class with a square-integrable envelope, so the empirical process indexed by it is asymptotically equicontinuous in the $L^2(P_a)$ distance \citep{vaart1996}. Consistency of the empirical median, the absence of an atom and $P_a\nu_a^2<\infty$ imply that the $L^2(P_a)$ distance between the indicators at the empirical and population medians tends to zero, so the remainder is $\op(1)$. Under condition (ii), the argument in the proof of Lemma~\ref{lem:quantile} applies: the error of $\widehat\lambda_{k,j}$ multiplies a centred sum that is $\Op(1)$.

If the treated proportion of fold $k$ is $p_k=p+O(m^{-1})$, the coefficients $1/p_k$ and $1/(1-p_k)$ of the transformed outcome differ from those at $p$ by $O(m^{-1})$, which changes the sums standardized by $m^{1/2}$ by $\op(1)$ under the moment conditions. Fold sizes that differ by a bounded number of units are covered by the same argument.

\section{Experiments}\label{app:experiments}

\subsection{Simulations in the model of the paper}\label{app:gatessim}

Unless stated otherwise, the simulations use the model of \S~\ref{sec:univariate}: one standard normal covariate, normal errors, half of each fold treated, and fold $k$ ranked at its median by the least-squares slope learned on the other folds.

\textit{Two folds.} Table~\ref{tab:finite} shows that the statistic is close to its limit with as few as 200 units. Its variance and kurtosis are close to the limiting values 1.405 and 2.489, normal critical values over-reject at every sample size, normal critical values scaled by the limiting standard deviation under-reject, and the critical value of $T_2$ holds the level. A uniform covariate gives the same picture.

\begin{table}[htbp]
\centering\footnotesize
\caption{Variance, kurtosis and rejection rates (\protect\%) of nominal 5\protect\% tests for the two-fold statistic with $n$ units}\label{tab:finite}
\begin{tabular}{rlccccc}
\toprule
$n$ & Covariate & Variance & Kurtosis & Normal & Normal, limiting SD & $T_2$ critical value\\
\midrule
200 & Normal & 1.426 & 2.52 & 8.9 & 4.0 & 5.3\\
1000 & Normal & 1.360 & 2.52 & 8.4 & 3.5 & 4.8\\
4000 & Normal & 1.404 & 2.49 & 8.9 & 3.6 & 4.8\\
1000 & Uniform & 1.500 & 2.40 & 9.6 & 3.5 & 5.2\\
\bottomrule
\end{tabular}
\end{table}

\textit{$K$ folds.} Table~\ref{tab:kfold} compares the statistic at $n=6000$ with its limit $T_K$ in variance, kurtosis and rejection rate; the two agree closely.

\begin{table}[htbp]
\centering\small
\caption{Variance, kurtosis and rejection rates (\protect\%) of nominal 5\protect\% tests for the $K$-fold statistic at $n=6000$ (finite) and for its limit $T_K$ (limit)}\label{tab:kfold}
\setlength{\tabcolsep}{4pt}
\begin{tabular}{rcccccccc}
\toprule
 & \multicolumn{2}{c}{Variance} & \multicolumn{2}{c}{Kurtosis} & \multicolumn{2}{c}{Normal} & \multicolumn{2}{c}{$T_K$ critical value}\\
\cmidrule(lr){2-3}\cmidrule(lr){4-5}\cmidrule(lr){6-7}\cmidrule(lr){8-9}
$K$ & finite & limit & finite & limit & finite & limit & value & finite\\
\midrule
2 & 1.383 & 1.405 & 2.46 & 2.49 & 8.4 & 8.8 & 2.21 & 4.8\\
5 & 1.635 & 1.613 & 2.34 & 2.35 & 11.2 & 11.0 & 2.32 & 5.1\\
10 & 1.901 & 1.885 & 2.44 & 2.43 & 15.2 & 15.1 & 2.55 & 5.2\\
\bottomrule
\end{tabular}
\end{table}

\textit{Variance corrections.} Table~\ref{tab:cvgates} complements Table~\ref{tab:references}. The column $r$ is the correlation between two fold estimates implied by Theorem~\ref{thm:counter}, $1/K$ is the correlation at which the correction of Nadeau and Bengio is exact, and the ratio is the expected corrected variance over the exact variance, from~\eqref{eq:nb}. The correction under-rejects where the ratio is clearly above one and over-rejects where it is below one. At $K=10$ the ratio is 1.01, yet the corrected test rejects 5.3\% of the time, because its rejection rate depends also on the randomness of the variance estimate and on the shape of the limit. With the exact variance, the limit rejects 3.7\%, 3.3\%, 3.7\%, 5.2\% and 8.0\% of the time for $K=2$, 5, 10, 20 and 50.

\begin{table}[htbp]
\centering\small
\caption{Rejection rates (\protect\%) of nominal 5\protect\% tests with variance corrections at $n=6000$. NB, correction of Nadeau and Bengio; $T_K$, critical value of the limit. Power is for $\tau(x)=8x/n^{1/2}$}\label{tab:cvgates}
\begin{tabular}{rccccccc}
\toprule
& \multicolumn{3}{c}{Correlation and ratio} & \multicolumn{2}{c}{Level} & \multicolumn{2}{c}{Power}\\
\cmidrule(lr){2-4}\cmidrule(lr){5-6}\cmidrule(lr){7-8}
$K$ & $r$ & $1/K$ & ratio & paired $t$ & $T_K$ & NB & $T_K$\\
\midrule
2 & 0.405 & 0.500 & 1.27 & 7.8 & 5.1 & 11.1 & 83.6\\
5 & 0.153 & 0.200 & 1.18 & 11.5 & 5.1 & 36.0 & 81.2\\
10 & 0.098 & 0.100 & 1.01 & 16.8 & 5.1 & 49.8 & 73.5\\
20 & 0.067 & 0.050 & 0.85 & 19.9 & 4.6 & 58.6 & 58.2\\
50 & 0.041 & 0.020 & 0.64 & 19.4 & 4.8 & 63.2 & 21.3\\
\bottomrule
\end{tabular}
\end{table}

\textit{Rejection regions.} In the limit, the symmetric 5\% region of $T_K$ has probability 2.5\% in each tail for $K=2$, 4.0\% in the left tail and 1.0\% in the right for $K=10$, and 4.95\% in the left tail for $K=50$, where the region is $|T|>4.01$ and the equal-tailed region is $T<-4.57$ or $T>2.24$. A signal shifts the statistic to the right, so the symmetric region loses power as $K$ grows. Table~\ref{tab:equaltail} shows this for $n=6000$ and $\tau(x)=ax/n^{1/2}$, both with the quantiles of $T_K$, which are unknown in practice, and with joint calibration using the covariance estimated in each data set and 999 draws.

\begin{table}[htbp]
\centering\small
\caption{Rejection rates (\protect\%) of the symmetric and equal-tailed 5\protect\% regions, in that order, for $\tau(x)=ax/n^{1/2}$}\label{tab:equaltail}
\setlength{\tabcolsep}{3pt}
\begin{tabular}{rcccccccc}
\toprule
& \multicolumn{4}{c}{Quantiles of $T_K$} & \multicolumn{4}{c}{Joint calibration}\\
\cmidrule(lr){2-5}\cmidrule(lr){6-9}
$K$ & $a=0$ & $a=2$ & $a=4$ & $a=8$ & $a=0$ & $a=2$ & $a=4$ & $a=8$\\
\midrule
2 & 5.1 / 5.1 & 9.7 / 9.7 & 26.8 / 26.9 & 83.2 / 83.2 & 5.0 / 5.1 & 10.0 / 9.9 & 26.4 / 26.6 & 83.0 / 82.9\\
5 & 5.0 / 5.0 & 8.2 / 9.5 & 24.2 / 26.9 & 80.9 / 82.5 & 5.0 / 5.0 & 8.5 / 9.3 & 23.8 / 26.4 & 80.2 / 82.4\\
10 & 5.0 / 4.7 & 6.2 / 8.7 & 17.8 / 25.7 & 74.3 / 83.5 & 5.1 / 4.9 & 6.3 / 8.7 & 17.6 / 25.7 & 74.2 / 83.0\\
20 & 5.1 / 5.0 & 4.5 / 9.8 & 9.1 / 26.0 & 59.7 / 83.2 & 4.9 / 5.0 & 4.5 / 9.7 & 8.9 / 25.7 & 58.9 / 82.8\\
50 & 4.4 / 4.7 & 3.1 / 8.5 & 1.6 / 25.7 & 20.9 / 82.2 & 4.5 / 4.8 & 3.2 / 8.4 & 1.8 / 25.7 & 21.6 / 81.9\\
\bottomrule
\end{tabular}
\end{table}

\textit{Three folds, three rankings and binary outcomes.} Three covariates are independent and uniform on $[0,1]$; given $X$, $Y(0)$ and $Y(1)$ are independent Bernoulli variables with means $0.2+0.2X_1$ and $0.3+0.2X_1$, so the conditional average effect is 0.1 everywhere while individual effects vary. Each fold has as many treated as control units, the candidate rankings are the median splits of the three covariates, and the training moments of the other two folds select a covariate and its sign. Table~\ref{tab:general} shows that joint calibration holds the level where normal critical values, which ignore the dependence between folds, do not.

\begin{table}[htbp]
\centering\small
\caption{Rejection rates (\protect\%) of nominal 5\protect\% tests with three folds, three candidate rankings and binary outcomes}\label{tab:general}
\begin{tabular}{rccc}
\toprule
$n$ & Joint, symmetric & Joint, equal tails & Normal\\
\midrule
900 & 4.9 & 4.9 & 13.2\\
3600 & 4.1 & 3.9 & 12.1\\
\bottomrule
\end{tabular}
\end{table}

\textit{Removing the non-normal component.} To illustrate the last paragraph of \S~\ref{sec:structure}, each fold is also evaluated through $U_k-\hat\Gamma Z_k$, with $\hat\Gamma$ estimated from the contributions of the units to the two moments, and the resulting statistic is divided by its estimated standard deviation. With $n=6000$ and 4000 data sets for each of $K=2$, 5 and 10, this statistic had variance between 0.98 and 1.02 and rejected 4.8--5.6\% of the time under the null hypothesis, as Proposition~\ref{prop:structure} predicts. Under $\tau(x)=4x/n^{1/2}$ and $8x/n^{1/2}$ it still rejected 4.8--5.3\% of the time, whereas the usual statistic rejected about 37\% and 89\% of the time with normal critical values.

\subsection{Reference distributions that refit the rankings, and further settings}\label{app:refit}

\textit{Design.} Each fold has a fixed number of treated units, and the ranking of fold $k$ is the sign of $\sum_{r\neq k}Z_r$ times the median split of the covariate; in the setting with five rankings it is the median split of the covariate with the largest absolute coordinate of $\sum_{r\neq k}Z_r$, with that coordinate's sign. All tests are two-sided at the 5\% level and use the same data sets. Joint calibration estimates $\Sigma$ by~\eqref{eq:designcov} and uses 999 draws.

\textit{Randomization test.} The treatment labels are redrawn within each fold, keeping its number of treated units, 499 times; for every draw the rankings are refitted and the usual standardized statistic is recomputed, and the symmetric and equal-tailed $p$-values are formed as in Algorithm~\ref{alg:calibration}. Under the sharp null hypothesis that $Y_i(1)=Y_i(0)$ for every unit, this test is exact.

\textit{Rank-transformed subsampling.} Subsamples of about $n/\log n$ units are formed as in Algorithm~2 of \citet{guo2025rts}: the units of each arm are permuted and cut into blocks, and matching blocks of the two arms form one subsample, giving 800 subsamples when $n=6000$. Each subsample is split into $K$ folds balanced by arm, and its fold statistics $t_k=\sign(\sum_{r\ne k}Z_r)U_k/\widehat\sigma_k$ are computed with the within-arm standard deviation $\widehat\sigma_k$ of the unit contributions. All fold statistics are ranked jointly and transformed to normal scores, the scores of each subsample are summed and divided by $K^{1/2}$, and the full-sample statistic of the same form is compared with these sums in absolute value.

\textit{Best linear predictor.} Fold $k$ regresses $\tilde Y$ on its centred covariate; the $t$-statistic of the slope, signed by the slope fitted on the other folds, is summed over folds and divided by $K^{1/2}$. Joint calibration draws only the training moments, because the evaluation moment of each fold equals its training moment.

\textit{Computing time.} Times are means per data set with $n=6000$ and include computing the statistic. A refit here is a closed-form slope, so the ratios understate the cost of refitting for learners that are expensive to fit.

\subsection{The Hillstrom analysis}\label{app:hillstrom}

Throughout \S\S~\ref{app:hillstrom} and~\ref{app:realtrials}, $S$ denotes the usual standardized statistic of \S~\ref{sec:usual}.

\textit{Data.} The experiment randomized 64,000 customers to three arms; we use the 42,613 customers of the men's e-mail and no-e-mail arms and treat the assignment as complete randomization of independent customers. A visit within two weeks was recorded for 10.6\% of control and 18.3\% of treated customers. Each arm is split at random into two folds, and both folds learn a positive slope, so both rank customers with higher past spending as the high-effect group.

\textit{Two folds.} The estimated correlation between the training and evaluation moments is 0.84 in one fold and 0.85 in the other. Joint calibration gives $p=0.065$ and a 5\% critical value of 2.22 for the absolute statistic. Rank-transformed subsampling gives 0.054--0.064 for three subsample sizes; here the subsamples are drawn independently within arms rather than as blocks of permutations. Treating the folds as independent gives 0.035. For comparison, the two single-split tests give 0.21 and 0.085 (Bonferroni combination 0.17), the paired $t$-test and the correction of Nadeau and Bengio, each with one degree of freedom, give 0.093 and 0.158, and a full-sample test of the linear moment $E[\{X-E(X)\}(\tilde Y-\tau)]=0$, which weights heterogeneity differently, gives 0.028.

\textit{Five and ten folds.} Table~\ref{tab:hillstromkfold} repeats the analysis with $K=5$ and 10, each fold ranked by the sign learned on the others; every fold is ranked in the same direction, and the training--evaluation correlations lie between 0.83 and 0.87. The upper quantile of the reference distribution stays near 2.23 while its left tail widens with $K$, as in the limit, and at $K=10$ the equal-tailed $p$-value is half the symmetric one.

\begin{table}[htbp]
\centering\small
\caption{The Hillstrom experiment: statistic $S$, $p$-values, quantiles of the reference distribution of joint calibration, and probability (\protect\%) of each tail of its symmetric 5\protect\% region}\label{tab:hillstromkfold}
\setlength{\tabcolsep}{3pt}
\begin{tabular}{rcccccccc}
\toprule
& & \multicolumn{3}{c}{$p$-value} & \multicolumn{2}{c}{Quantiles} & \multicolumn{2}{c}{Symmetric region}\\
\cmidrule(lr){3-5}\cmidrule(lr){6-7}\cmidrule(lr){8-9}
$K$ & $S$ & normal & joint, symmetric & joint, equal tails & 2.5\% & 97.5\% & left & right\\
\midrule
2 & 2.11 & 0.035 & 0.066 & 0.065 & $-2.22$ & 2.22 & 2.5 & 2.5\\
5 & 1.98 & 0.047 & 0.111 & 0.092 & $-2.43$ & 2.24 & 3.1 & 1.9\\
10 & 2.19 & 0.029 & 0.111 & 0.057 & $-2.84$ & 2.24 & 4.1 & 0.9\\
\bottomrule
\end{tabular}
\end{table}

\textit{Redrawn labels.} To check the two reference distributions on these data, we keep the covariate and outcome of every customer, redraw the e-mail labels by complete randomization with the observed arm sizes, so that the labels have no effect on any customer, and repeat the analysis on 2000 data sets for each $K$. Table~\ref{tab:hillstromrerand} gives the rejection rates. The variance of the statistic, 1.46, 1.67 and 1.97, is close to the values 1.46, 1.70 and 2.00 that~\eqref{eq:varTK} gives for the observed correlation 0.85.

\begin{table}[htbp]
\centering\small
\caption{Rejection rates (\protect\%) on the Hillstrom data under redrawn treatment labels}\label{tab:hillstromrerand}
\begin{tabular}{rcccc}
\toprule
$K$ & Normal & Joint, symmetric & Joint, equal tails & Variance of $S$\\
\midrule
2 & 8.8 & 4.9 & 4.9 & 1.46\\
5 & 11.0 & 5.0 & 5.0 & 1.67\\
10 & 16.8 & 5.0 & 4.5 & 1.97\\
\bottomrule
\end{tabular}
\end{table}

\subsection{Experiments with redrawn treatment labels}\label{app:realtrials}

\textit{Data and labels.} From ACTG 175 \citep{actgdata,hammer1996} we use the 1054 patients randomized to zidovudine or to zidovudine plus didanosine, with 16 baseline covariates and the change in CD4 count at week 20 as outcome; the trial was stratified by prior antiretroviral therapy, so labels are redrawn and covariances computed within strata. From the Criteo uplift data \citep{criteodata,diemert2018} we use a uniform sample of 20,000 units with twelve features and a website visit as outcome. In every data set the covariates and outcomes stay fixed and the labels are redrawn by complete randomization with the original numbers in each stratum and arm, so that the labels have no effect on any unit.

\textit{One covariate against all covariates.} Table~\ref{tab:rtonedim} compares a least-squares ranking on one covariate with the same ranking on all covariates, under leave-one-fold-out with folds balanced within stratum and arm.

\begin{table}[htbp]
\centering\small
\caption{Variance, excess kurtosis and rejection rate (\protect\%) of the nominal 5\protect\% test standardized by the exact standard deviation, with redrawn treatment labels}\label{tab:rtonedim}
\begin{tabular}{llcccccc}
\toprule
& & \multicolumn{3}{c}{$K=2$} & \multicolumn{3}{c}{$K=5$}\\
\cmidrule(lr){3-5}\cmidrule(lr){6-8}
& Covariates & Var & Ex.\ kurt. & Exact SD & Var & Ex.\ kurt. & Exact SD\\
\midrule
ACTG & CD4 & 1.36 & $-$0.26 & 4.6 & 1.53 & $-$0.51 & 3.7\\
ACTG & weight & 1.28 & $-$0.34 & 4.5 & 1.44 & $-$0.43 & 4.3\\
ACTG & all 16 & 1.53 & $-$0.09 & 4.6 & 1.58 & $-$0.13 & 4.6\\
Criteo & $f_0$ & 1.49 & $-$0.67 & 3.1 & 1.69 & $-$0.77 & 2.8\\
Criteo & $f_6$ & 1.41 & $-$0.63 & 3.5 & 1.58 & $-$0.64 & 2.9\\
Criteo & all 12 & 1.43 & $-$0.22 & 4.6 & 1.47 & $-$0.07 & 5.1\\
\bottomrule
\end{tabular}
\end{table}

\textit{A finite set of rankings.} Median splits of continuous covariates, with their sign, are selected by the largest absolute coordinate of the summed training moments: age, weight, days of prior therapy, and CD4 and CD8 counts on ACTG ($J=5$), and the six Criteo features whose medians lie strictly inside the range of their values ($J=6$). This setting falls under Theorem~\ref{thm:calibration}. Table~\ref{tab:rtdict} also reports calibration of the sum $S_n=K^{-1/2}\sum_k\hat\omega_k^\T U_{k,n}$, without the standardization by $\hat v$ in \eqref{eq:Tn}, which differs from $T_n$ when the candidate rankings have different variances.

\begin{table}[htbp]
\centering\footnotesize
\caption{Rejection rates (\protect\%) of nominal 5\protect\% tests with a finite set of candidate rankings and redrawn treatment labels}\label{tab:rtdict}
\setlength{\tabcolsep}{4pt}
\begin{tabular}{lrrccccc}
\toprule
& & & & \multicolumn{2}{c}{Joint, $T_n$} & \multicolumn{2}{c}{Joint, $S_n$}\\
\cmidrule(lr){5-6}\cmidrule(lr){7-8}
& $J$ & $K$ & Normal & Symmetric & Equal tails & Symmetric & Equal tails\\
\midrule
ACTG & 5 & 2 & 8.6 & 5.7 & 5.6 & 5.7 & 5.6\\
ACTG & 5 & 5 & 11.5 & 5.7 & 5.7 & 5.5 & 5.7\\
Criteo & 6 & 2 & 11.0 & 5.7 & 5.9 & 5.3 & 5.5\\
Criteo & 6 & 5 & 13.2 & 4.5 & 4.6 & 4.5 & 4.8\\
\bottomrule
\end{tabular}
\end{table}

\textit{A correction for many covariates.} For the correction of \S~\ref{sec:repairs}, fold $k$ is ranked by least squares of $\tilde Y$ on the covariates, fitted on the other folds, and split at the median of its score. Let $\hat\nu_i=\tilde Y_i-\hat\tau$, and let $t_i=(X_i-\bar X)\hat\nu_i$ and $e_{k,i}=\hat g_k(X_i)\hat\nu_i$ be the contributions of unit $i$ to the training moment and to the evaluation moment under $\hat g_k$, the median split of the score of fold $k$. Let $\hat C$ be the covariance matrix of $(t_i,e_{k,i})$ over all units, computed within each combination of stratum and arm. Then $\hat\rho_k^2=\hat c_k^\T\hat C_{tt}^{-1}\hat c_k/\hat C_{ee}$, where $\hat c_k$ is the covariance of $t_i$ and $e_{k,i}$, and the corrected statistic is $S(1+\bar\rho^{\,2})^{-1/2}$ with $\bar\rho^{\,2}$ the average of $\hat\rho_k^2$ over folds. In Table~\ref{tab:rtcorrect}, with all covariates the variance of $S$ lies within 10\% of $1+\bar\rho^{\,2}$. With one covariate $1+\bar\rho^{\,2}$ exceeds it by up to 16\% at $K=2$, as the variance excess $2\rho^2/\pi$ of Theorem~\ref{thm:counter} predicts, and the light tails make the corrected test conservative at $K=5$ as well.

\begin{table}[htbp]
\centering\small
\caption{The variance correction of \S~\ref{sec:repairs} with redrawn treatment labels: variance of $S$, its estimate $1+\bar\rho^{\,2}$, and rejection rates (\%) of nominal 5\% tests}\label{tab:rtcorrect}
\begin{tabular}{llrccccc}
\toprule
Trial & Covariates & $K$ & Var & $1+\bar\rho^{\,2}$ & Normal & Exact SD & Corrected\\
\midrule
ACTG & all 16 & 2 & 1.54 & 1.62 & 11.3 & 4.8 & 4.4\\
ACTG & all 16 & 5 & 1.64 & 1.61 & 13.1 & 5.0 & 5.2\\
ACTG & CD4 & 2 & 1.30 & 1.45 & 8.5 & 3.9 & 2.8\\
ACTG & CD4 & 5 & 1.39 & 1.45 & 9.5 & 4.0 & 3.8\\
Criteo & all 12 & 2 & 1.52 & 1.65 & 10.7 & 4.4 & 3.2\\
Criteo & all 12 & 5 & 1.70 & 1.65 & 13.4 & 5.0 & 5.2\\
Criteo & $f_0$ & 2 & 1.55 & 1.80 & 9.3 & 2.8 & 1.6\\
Criteo & $f_0$ & 5 & 1.72 & 1.80 & 10.5 & 2.8 & 2.1\\
\bottomrule
\end{tabular}
\end{table}

\textit{Observed labels.} With the observed labels, the test based on the finite set of rankings shows no heterogeneity on ACTG~175, whichever reference distribution is used. On Criteo with $K=5$ the normal $p$-value is 0.0025, and joint calibration and a randomization reference both give 0.011 for $T_n$; calibrating $S_n$ instead gives 0.083, because the selected ranking has a small variance and the reference distribution of $S_n$ is dominated by rankings with larger ones.

\end{document}